\documentclass[useAMS,usenatbib]{mn2e}

\usepackage{epsfig, natbib, soul}

\usepackage{color}

\usepackage{amsmath}    
\usepackage{mathtools}  
\usepackage{amssymb}    
\usepackage{verbatim}   
\usepackage{color}      
\usepackage{hyperref}   
\usepackage{multicol}  %
\usepackage{multirow}
\usepackage{ifthen}  

\usepackage{graphicx}
\usepackage{subcaption}

\usepackage{lipsum}       
\usepackage{changepage}   
\usepackage{enumitem}

\newcounter{CDMDone}
\def\CDM{\ifthenelse{\equal{\arabic{CDMDone}}{0}}{cold dark matter (CDM)\setcounter{CDMDone}{1}}{CDM}}

\newcounter{RMDone}
\def\RM{\ifthenelse{\equal{\arabic{RMDone}}{0}}{redMaPPer \setcounter{RMDone}{1}}{redMaPPer }}

\newcounter{haloDone}
\def\hc{\ifthenelse{\equal{\arabic{haloDone}}{0}}{ halo-cluster \setcounter{haloDone}{1}}{ halo-cluster }}

\newcounter{clusterDone}
\def\ch{\ifthenelse{\equal{\arabic{clusterDone}}{0}}{ cluster-halo \setcounter{clusterDone}{1}}{ cluster-halo }}

\newcommand{\kms}{{\rm \, km~s}\ensuremath{^{-1}}}

\newcommand{\hinv}{\ensuremath{\, h^{-1}}}%
\newcommand{\msol}{\ensuremath{\, {\rm M}_\odot}}         
\newcommand{\kpc}{\ensuremath{\, {\rm kpc}}}         
\newcommand{\mpc}{\ensuremath{\, {\rm Mpc}}}         
\newcommand{\gpc}{\ensuremath{\, {\rm Gpc}}}%
\newcommand{\gyr}{\ensuremath{\, {\rm Gyr}}}%

\newcommand{\apj}{ApJ}
\newcommand{\apjs}{ApJS}
\newcommand{\apjl}{ApJL}
\newcommand{\aap}{A{\&}A}

\newcommand{\mnras}{MNRAS}
\newcommand{\aj}{AJ}
\newcommand{\araa}{ARAA}
\newcommand{\pasp}{PASP}

\newcommand{\pmem}{P_{\rm mem}}         
\newcommand{\pmemalpha}{P_{\rm mem, \alpha}}         
        
\newcommand{\phaloi}{P_{{\rm halo},i} }        
         
\newcommand{\Rtwoh}{R_{\rm 200c}}        
\newcommand{\rtwoh}{R_{\rm 200c}}        
\newcommand{\mtwoh}{M_{\rm 200c}}

\newcommand{\avg}[1]{\langle #1 \rangle}

\begin{document}

\title[Galaxy Cluster Mass from Stacked Spectroscopy]{Galaxy Cluster Mass Estimation from Stacked Spectroscopic Analysis}
\author[A. Farahi {\sl et al.}]{Arya Farahi$^{1}$\thanks{E-mail: aryaf@umich.edu}, August E. Evrard$^{1,2}$,
Eduardo Rozo$^{3,4}$,
Eli S. Rykoff$^{4}$,
\newauthor 
Risa H. Wechsler$^{4,5,6}$ \\
$^{1}$Department of Physics and Michigan Center for Theoretical Physics, University of Michigan, Ann Arbor, MI 48109 USA\\
$^{2}$Department of Astronomy, University of Michigan, Ann Arbor, MI 48109 USA\\
$^{3}$Department of Physics, University of Arizona, 1118 E 4th St, Tucson, AZ 85721, USA\\
$^{4}$SLAC National Accelerator Laboratory, Menlo Park, CA 94025, USA\\
$^{5}$Kavli Institute for Particle Astrophysics and Cosmology, P.O. Box 2450, Stanford, CA 94305\\
$^{6}$Department of Physics, Stanford University, 382 Via Pueblo Mall, Stanford, CA 94305
}

\pagerange{\pageref{firstpage}--\pageref{lastpage}} \pubyear{2016}

\maketitle

\label{firstpage}

\begin{abstract}
We use simulated galaxy surveys  
to study: i) how galaxy membership in redMaPPer clusters maps to the underlying halo population, and ii) the accuracy of a mean dynamical cluster mass, $M_\sigma(\lambda)$, derived from stacked pairwise spectroscopy of clusters with richness $\lambda$.  
Using $\sim\! 130,000$ galaxy pairs patterned 
after the SDSS redMaPPer cluster sample study of \citet[][ RMIV]{Rozo:2014}, we show 
that the pairwise velocity PDF of central--satellite pairs with $m_i < 19$ in the
simulation matches the form seen in RMIV.  Through joint membership matching, we deconstruct the 
main Gaussian velocity component into its halo contributions, finding that the top-ranked halo contributes $\sim 60\%$ of the stacked signal. 
The halo mass scale inferred by applying the virial scaling of \cite{Evrard:2008} to the velocity
normalization matches, to within a few percent, the log-mean halo mass derived through galaxy membership matching.  We apply this approach, along with mis-centering and galaxy velocity bias corrections, to estimate the log-mean matched halo mass at $z=0.2$ of SDSS redMaPPer clusters.  Employing the velocity bias constraints of \citet{GuoII:2015}, we find $\avg{\ln(M_{200c})|\lambda} = \ln(M_{30}) + \alpha_m \ln(\lambda/30)$ with $M_{30} = 1.56 \pm 0.35 \times 10^{14} \msol$ and $\alpha_m = 1.31 \pm 0.06_{stat} \pm 0.13_{sys}$. 
Systematic uncertainty in the velocity bias of satellite galaxies overwhelmingly dominates the error budget.
\end{abstract}



\section{Introduction}\label{sec:Introduction}

The most massive dark matter halos to emerge in the universe host clusters of galaxies.  
Ongoing and near-future cosmological surveys are dedicated to identifying clusters for the purpose of studying cosmology and fundamental physics through spatiotemporal counts and other statistical properties of the cluster population \citep[][]{Allen:2011}.   
The largest cluster samples are identified using photometric data, through color-based \citep{GladdersYee:2005, Koester:2007, Dong:2008, Murphy:2012, Oguri:2014, Stanford:2014, Bleem:2015, Licitra:2016} or photometric redshift-based \citep{Milkeraitis:2010, Durret:2011, Soares-Santos:2011} algorithms. 

Predicting such cluster counts for a given cosmology requires convolving the halo mass function (spatial number density as a function of mass and redshift) with a likelihood function linking observable cluster properties to total halo mass. As a result, the true halo mass of clusters is a crucial element in the methodology of cluster count cosmology.  

Because photometric data provides only coarse resolution in redshift, projection of galaxies along the line of sight to a massive halo limits the ability of cluster-finding algorithms to uniquely identify the galaxies that are members of a particular massive halo.  Spectroscopic data provides improved distance and mass estimators for group and cluster selection \citep[e.g.,][]{Robotham:2011}, but projection and mis-centering still pose challenges for these methods \citep[see e.g.][and references therein]{Duarte:2014}.  

These sources of confusion are fundamentally rooted in the fact that clusters and halos are identified in different spaces: sky-redshift or sky-color space for clusters and 3D real space or 6D phase space for halos.
Peculiar velocities can blend distinct halos in real space into a single structure in redshift space \citep[{\sl e.g.},][]{vandenBosch:2004, Biviano:2006, Wojtak:2007, Saro:2013, Duarte:2014}.  In addition, the fact that high mass halos in cold dark matter cosmologies are dynamically evolving at late times means that substructure and mergers can create complex, transient phase-space structure.  In simulations, this complexity can confuse assignment of subhalos hosting galaxies to their parent halos \citep{Knebe:2011}.  

In practice, assigning galaxies as members of either clusters or halos is a matter of convention, defined by application of specific, algorithm-dependent rules to galaxy samples.  Regardless of the particulars, the joint likelihood, $P_{\alpha,i}(k)$ that a galaxy, $k$, is a member of both cluster $\alpha$ and halo $i$ offers a means to map from one space to the other \citep{Gerke:2005}.  

The total galaxy content, or \emph{richness}, of a cluster can then be considered as a sum of partial contributions from halos closely aligned along a common sightline.  In this work, we apply such a membership-matching approach in simulations to build a network linking clusters to halos, with network edges weighted by fractional cluster membership.  

We investigate the membership properties of the redMaPPer cluster finding algorithm \citep{Rykoff:2014}.  The method, which identifies clusters through their red sequence galaxy population, outputs background-corrected membership probabilities \citep{Rozo:2009, Rykoff:2012} to each galaxy in a cluster as well as central galaxy probabilities for up to four cluster members.  The method is designed to make optimal use of data from large, multi-color photometric surveys such as the Sloan Digital Sky Survey \citep[SDSS,][]{SDSS-York:2000} and the Dark Energy Survey \citep[DES,][]{DES-Flaugher:2005}.   
The SDSS redMaPPer cluster catalog \citep{Rykoff:2014} has been extensively studied with multiwavelength data, including comparisons to existing X-ray and Planck satellite Sunyaev-Zel'dovich catalogs \citep{Sadibekova:2014, RozoRykoff:2014, Rozo:2015, PlanckXXVII:2015}.

The latest study in the redMaPPer series uses stacked spectroscopic analysis of cluster member pairwise velocities to investigate photometrically assigned membership probabilities \citep[][hereafter RMIV]{Rozo:2014}.  In that work, very good agreement was found between spectroscopic and photometric definitions of cluster membership after a small number of modest corrections for blue cluster members, correlated line-of-sight structure, and photometric noise.  

Using only SDSS data, the RMIV study could not study membership from the perspective of the underlying halo population.  
Instead, spectroscopic members are defined in velocity space using an assumed Gaussian form for the pairwise velocity 
probability density function (PDF) of central and satellite cluster members. In this work, we use simulations to link spectroscopic cluster members to the underlying halo population, leading to an estimate of the log-mean matched halo mass. 

In \S\ref{sec:simulation}, we apply the redMaPPer algorithm to a 10,000 deg$^2$ synthetic photometric galaxy catalog derived from lightcone outputs of N-body simulations.  We then employ a membership-based matching algorithm, described in \S\ref{sec:matching}, to build bipartite graphs\footnote{A bipartite graph links two disjoint sets of nodes, {\sl U} and {\sl V}, with edges, each of which connects a node in {\sl U} with one in {\sl V}.  In our case {\sl U} is the set of clusters and {\sl V} the set of halos.}
in which each cluster links to a set of halos ranked by their fractional member contribution to that cluster, a measure we term membership {\sl strength}. This method is used to deconstruct the stacked pairwise velocity distribution of central-satellite galaxies in \S\ref{sec:Halos-spectroscopic}.   

In \S\ref{sec:massCalibration}, we apply the N-body simulation-based virial scaling of \citet{Evrard:2008} to estimate the total mass at fixed cluster richness from the velocity dispersion model of \S\ref{sec:Halos-spectroscopic}.  We show that this dynamical mass recovers the log-mean mass of halos matched by cluster membership to better than one percent. Confounding effects of mis-centering and velocity bias are then discussed.  Using current estimates for the magnitudes of these sources of systematic error, in \S\ref{sec:RMIVmass} we estimate the halo mass scale of the RMIV sample using their stacked velocity dispersion measurements.  
Our results are summarized in \S\ref{sec:conclusion}.

Unless otherwise noted, our convention for the mass of a halo is $M_{200c}$,  the mass contained within a spherical region encompassing a mean density equal to $200$ times the critical density of the universe, $\rho_c(z)$. We also test the robustness of our results to the details of the synthetic galaxy population by implementing our analysis on an independent, higher-resolution simulation, populated with a different galaxy prescription.  Appendix \ref{app:Bolshoi} summarizes these results.


\section{Simulation samples and synthetic cluster catalog} \label{sec:simulation}

We employ N-body simulations produced with a lightweight version of the Gadget code developed for the Millennium Simulation \citep{Springel:2005}.  Three simulations, of $1.05$, $2.6$ and $4.0 \hinv \gpc$ volumes, are used to produce a sky survey realization covering 10,000 deg$^2$ that resolves all halos above $10^{13} \msol$ within $z \le 2$.   We refer to this suite of runs as the Aardvark simulation.  

The resultant sky catalog is built by concatenating continuous lightcone output segments from the three different N-body volumes using the method described in \citet{Evrard:2002}.  The smallest volume maps $z < 0.35$, the intermediate maps $0.35 \le z < 1.1$ and the largest volume covers $1.1 \le z < 2$.  The simulations employ $2048^3$ particles, except for the $1.0 \hinv \gpc$ volume which uses $1400^3$, and corresponding particle masses are $0.27$, $1.3$ and  $4.8 \times 10^{11} \hinv \msol$.   The Aardvark suite assumes a $\Lambda$CDM cosmology with cosmological parameters: $\Omega_m = 0.23$, $\Omega_{\Lambda} = 0.77$, $\Omega_{b} = 0.047$, $\sigma_8 = 0.83$, $h = 0.73$, and $n_s = 1.0$.  The Rockstar algorithm is used for halo finding \citep{Behroozi:2013}.  

\subsection{Galaxy population and halo membership}

Galaxy properties are assigned to particles using the ADDGALS algorithm \citep{addgals1, addgals2,Chang2015,addgals3}.  The algorithm is empirical, using the observed $r$-band luminosity function and trend of galaxy color with local environment as input.  The method assigns central galaxies to resolved halos, but satellites as well as centrals in unresolved halos are assigned to dark matter particles in a probabilistic manner weighted by a local dark matter density estimate.  This density assignment scheme is tuned to match the clustering properties of a sub-halo assignment matching (SHAM) approach applied to a $400 \hinv\mpc$ simulation using $2048^3$ particles.

Central galaxies are placed at the center of resolved halos and assigned a velocity at rest relative to the halo's mean dark matter velocity within $\rtwoh$.  We explore the issue of non-zero central galaxy velocities in the analysis below.  
All other galaxies are assigned the positions and velocities of the corresponding particles to which they are assigned.  Note that no particle can host more than one galaxy.   The velocity assignment implies that the velocity dispersion of central--satellite pairs is expected to follow the same scaling with halo mass as that identified in the simulation ensemble of \citet{Evrard:2008}.  

Regarding halo membership, our convention is that a galaxy, $n$, is assigned to one and only one halo.  Thus, if galaxy $n$ is assigned to halo $j$, then the probability that galaxy $n$
belongs to halo $i$ is $\phaloi(n) = \delta_{ij}$.  A spherical region of radius $\Rtwoh$ is used when defining halo membership.  
This region approximately defines the hydrostatic region of massive halos but it does not extend to the 
outer caustic, or backsplash, edge which contains a mix of infalling and outgoing material \citep{Busha:2005, Cuesta:2008, More:2015}. 
We note that $\rtwoh$ is similar in scale to the search radius used by the \RM cluster finding algorithm. In regions where two or more halos spatially overlap, the galaxy is assigned to the nearest halo.  
In the ADDGALS algorithm, galaxies can reside outside of a resolved N-body halo; $13\%$ of $m_i < 19$ galaxies reside beyond $\rtwoh$ of a resolved halo. 

While not strictly a halo occupation distribution (HOD) method, ADDGALS produces an effective HOD for which intrinsic richness scales as a power law with halo mass.  At low redshift, $\lambda_{\rm int}$, defined as the number of galaxies with $M_r - 5 \log h \le  -19$ within $\rtwoh$, scales with halo mass in a sub-linear fashion, $\lambda_{\rm int} \propto M^\alpha$ with $\alpha \sim 0.8$.  To test the robustness of our conclusions to the intrinsic HOD structure of massive halos, we repeat the analysis on the galaxy catalogs of \citet{Hearin:2013} extracted from the Bolshoi simulation, which have a slightly steeper slope, $\alpha \sim 1.0$, and smaller intrinsic scatter in $\lambda_{\rm int}$ compared to the Aardvark galaxy catalog.  Further details are provided in Appendix~\ref{app:Bolshoi}.

\begin{figure}
   \centering
   \includegraphics[width=0.45\textwidth]{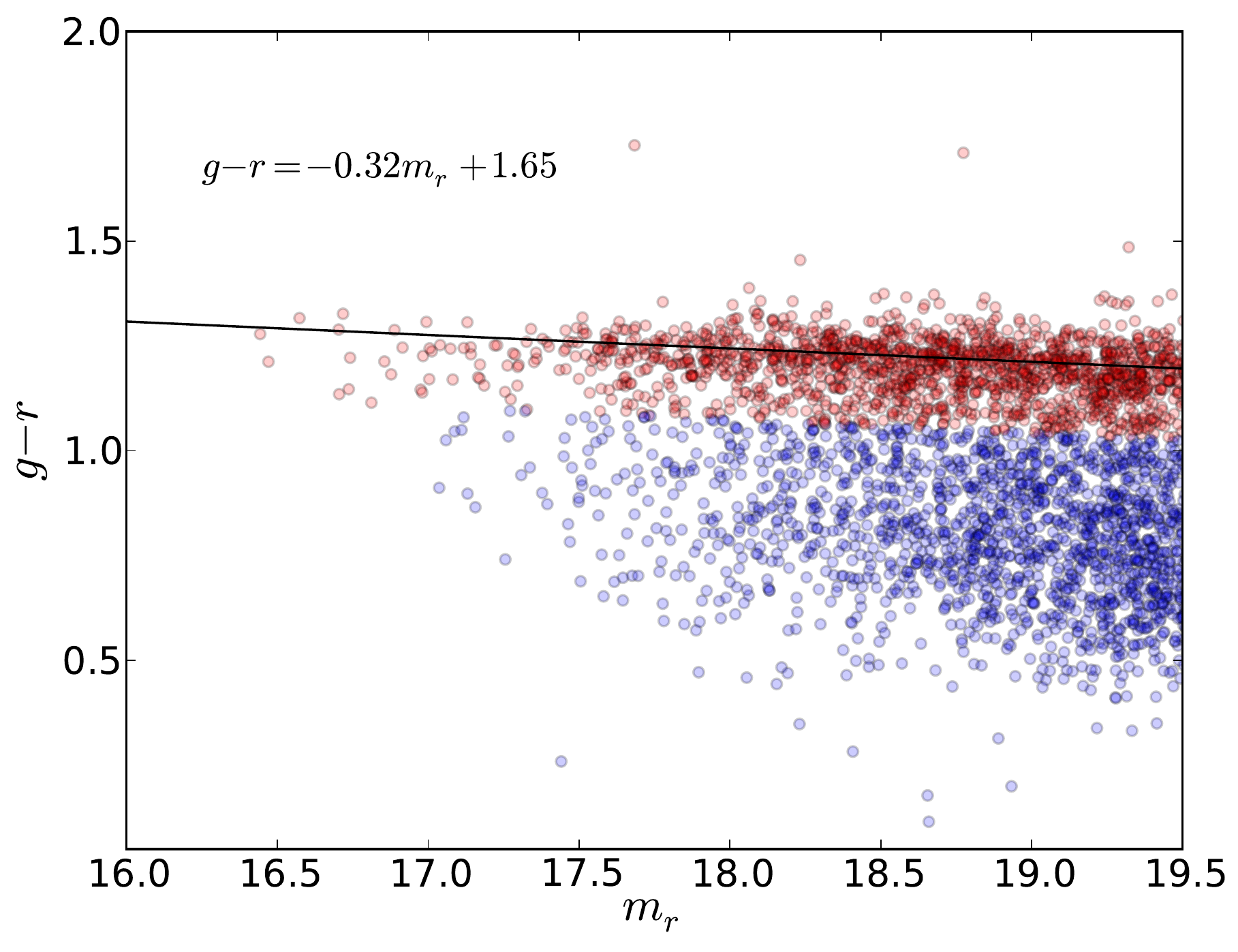}
 \caption{Color--magnitude diagram for Aardvark simulation galaxies occupying halos of mass $\mtwoh > 10^{14} \hinv\msol$ in the redshift interval $0.19 < z < 0.21$.  The line indicates the red sequence ridge-line, $g-r = 1.65 - 0.32 \, m_r$; $78\%$ of galaxies brighter than $m_i = 19$ lie within 0.2 mag of this ridge-line.  } 
   \label{fig:colorMag} 
\end{figure}

The \RM algorithm assumes that red galaxies are the prominent population occupying high mass halos.  In 
Figure~\ref{fig:colorMag}, we show the distribution of $g-r$ color as a function of $r$-band magnitude, $m_r$, for Aardvark galaxies in halos of mass $\mtwoh > 10^{14} \hinv\msol$, and in the narrow redshift interval, $0.19 < z < 0.21$.   A red sequence is evident, containing $78\%$ of galaxies brighter than $19^{\rm th}$ magnitude. The line shows the ridge-line approximate red sequence population. The slope and intercept are consistent with those found in SDSS analysis of \citet[][see their Fig.~11]{Hao:2009} for the same redshift range. 
 
While the ADDGALS method uses a local dark matter density to assign galaxy luminosity to particles, the smoothing scale employed to calculate the local density leaves the inner $\sim 100 \kpc$ of high mass halos relatively devoid of galaxies other than the central.  As a result of this and possibly other factors, the frequency of mis-centered clusters is larger in the Aardvark \RM cluster catalog than in the observed SDSS sample.  We therefore work with two different cluster samples, consisting of the correctly centered subset (denoted CEN) as well as the full set of identified \RM clusters (ALL). 
The exact definition of these two samples is given in \S\ref{sec:specSamples}.

\subsection{Cluster finding with redMaPPer} \label{sec:redMaPPer-cluster-finder}

Cluster finding methods that use only optical photometry fall into two main categories based on whether the method uses colors directly or photometric redshifts derived from those colors.  The \RM algorithm is in the former category; it uses colors, along with training spectroscopy, to track the multi-band location of the red sequence as a function of redshift \citep{Rykoff:2014}.  
We note that \RM\ is continuously updated, so there is no unique \RM\ catalog.  Here, we rely on the \RM\ v5.10 SDSS catalog, as this constitutes the most recently
publicly available version.

The \RM cluster finder is a matched filter algorithm with components that characterize the luminosity function, red-sequence color, and projected number density of cluster galaxies.  Writing the projected galaxy distribution in sky-magnitude space as a sum of cluster members and a locally-uniform background component, the algorithm works iteratively to eventually tag each galaxy in the vicinity of a cluster, $\alpha$, with a probability, $\pmemalpha$ of being a member of that cluster.   The richness, $\lambda$, is defined as the sum of the membership probabilities over the set, $G_\alpha$, of all member galaxies
\begin{equation} 
  \lambda_{\alpha} = \sum\limits_{n \in G_{\alpha}} \pmemalpha(n) .
\label{eq:lambda}
\end{equation} 

\begin{figure}
   \centering
   \includegraphics[width=0.45\textwidth]{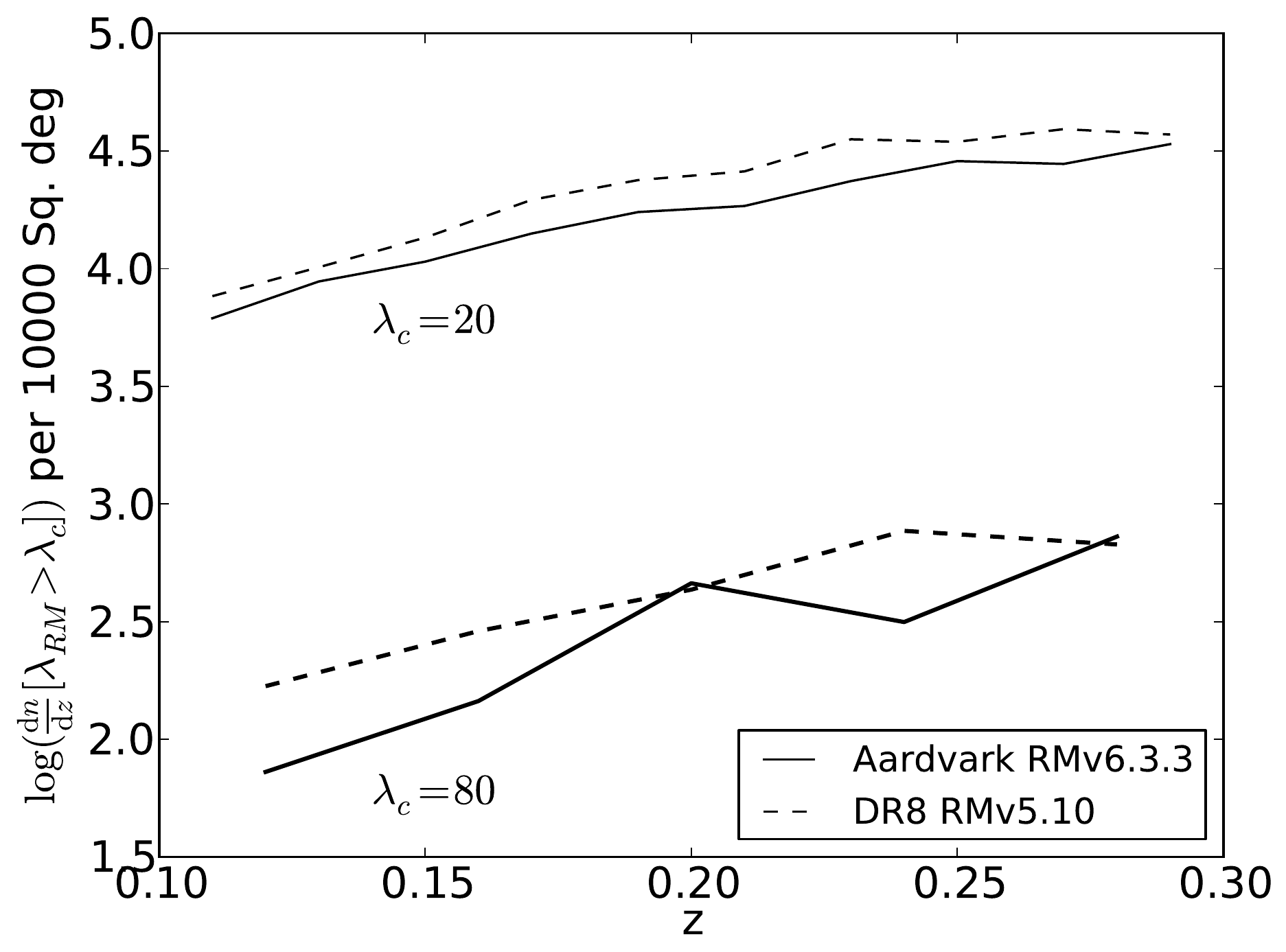}
 \caption{Differential sky number counts per $10,000$ square degree of clusters with richness, $\lambda > 20$ (thin lines) and $80$ (bold lines) are shown for the Aardvark simulated galaxy catalog run with RMv6.3.3 (solid) and SDSS DR8 run with RMv5.10 \citep[dashed,][]{Rozo:2014} samples. }
   \label{fig:N-z-lam}
\end{figure}

The \RM algorithm applied to the Aardvark galaxy sample yields 3927 clusters with $\lambda > 20$ and redshift of $[0.1-0.3]$ over 10,400 square degrees.  By comparison, there are 4522 
clusters in the \RM v5.10 DR8 cluster sample. 
Figure~\ref{fig:N-z-lam} shows differential sky number counts, ${\rm d}n/{\rm d}z$, in units of number per $10,000$ square degrees, for clusters with $\lambda > 20$ (upper lines) and $80$ (lower lines) in the Aardvark and SDSS DR8 samples.

The number of clusters with $\lambda > 20$ in our simulation is lower by $\sim 24 \%$ relative to the SDSS DR8 catalogs. This suppression may partly reflect the underpopulation of the inner $\sim 150 \kpc$ regions of the most massive simulated halos, which suppresses the membership probability PDF at high $\pmem$ values for cluster members.  In addition, the lower central galaxy density of massive Aardvark halos makes it more difficult for \RM\ to center clusters correctly.  We note that the simulation matches well the observed trend of increasing counts with redshift. 
Finally, the difference may reflect differences in the underlying cosmological parameters. The Aardvark simulation has a smaller dark matter density ($\Omega_m=0.23$) than most current observational constraints, which implies a lower space density at fixed halo mass.  The small difference in overall counts does not influence the spectroscopic analysis below.

\subsection{Cluster and Spectroscopic samples } \label{sec:specSamples}

\begin{table*}
\centering
\caption{Aardvark cluster samples, including the number of redMaPPer clusters, $N_{\rm cl}$, the number of galaxies in the spectroscopic samples, $N_{\rm spec}$, and the number of spectroscopic, central-satellite pairs, $N_{\rm pair}$. 
}
\begin{tabular}{ccccl} 
Name & $N_{\rm cl}$  & $N_{\rm spec}$ & $N_{\rm pair}$ & Sample description  \\
\hline
ALL  & 3927  & 134464 & 130537 & full sample with $\lambda > 20$  \\
CEN & 2294  & 78794  & 76500 & correctly centered subsample of ALL \\
\hline
\end{tabular}
\label{tab:sampleDefs}
\end{table*}

The full \RM cluster catalogs for both observations and simulated galaxy catalogs consist of clusters with $\lambda > 20$ in the redshift range $z= 0.1$ to $0.3$.

To evaluate the sensitivity of our analysis to mis-centering, we identify a correctly centered sub-sample of simulated clusters, 
those for which the central cluster galaxy is also the central galaxy of the top-ranked, matched halo.  Throughout this work, we refer to this correctly centered sub-sample as CEN, and denote the full simulated cluster sample as ALL.

Our spectroscopic membership study is limited to  cluster member galaxies with $m_i < 19$.  
The limit of $m_i < 19$ is a compromise value lying between the SDSS and GAMA limits used by RMIV. Because satellite galaxies in halos trace the dark matter kinematics by construction, our results are not strongly sensitive to the choice of magnitude limit.  

Table \ref{tab:sampleDefs} summarizes the number of clusters, number of galaxies, and number of central--satellite galaxy pairs in the simulation samples used below.

\section{Cluster--Halo membership matching} \label{sec:matching}

To match \RM clusters to halos, we build a bipartite network between clusters and halos with edges weighted by joint cluster--halo membership.
The network is built using all photometric \RM members of the cluster.  Edges are weighted by the membership {\bf strength} between cluster $\alpha$ and halo $i$, defined as 
\begin{equation} 
S_{\alpha,i}= \frac{1}{\lambda_{\alpha}} \sum\limits_{n \in G_{\alpha}} P_{mem,\alpha}(n)P_{halo,i}(n) 
\label{eq:strength}
\end{equation}
where $G_{\alpha} \equiv \{ID\}_{\alpha}$ is the list of galaxy ID's associated with cluster $\alpha$, $P_{halo,i}(n)$ is a boolean set to 1 if galaxy $n$ is a member of halo $i$, as described in \S~\ref{sec:simulation}.  The strength, normalized to lie between $0$ and $1$, gives the fraction of the total membership of cluster $\alpha$ contributed by halo $i$.  

Recall that $\lambda_\alpha$ is the cluster richness defined in Equation~(\ref{eq:lambda}).  In essence, the measured optical richness of a cluster can be expressed as a series of decreasing halo contributions
\begin{equation} 
\lambda_{\alpha} =  \sum_{r=1}^N \, S_{\alpha,i(r)}  ,
\end{equation}
where the halo list, $i(r)$, is rank ordered such that 
$S_{\alpha,i(1)} \ge S_{\alpha,i(2)} \ge ... S_{\alpha,i(N)} $.  The matched halo of a cluster is defined as the halo with the highest strength; we use the terms ``matched halo'' and ``top-ranked halo'' interchangeably throughout this work.  The mapping is not exclusive; two clusters can be mapped to one halo. In practice this happens infrequently. Out of $3927$ \RM clusters of redshift $0.1$ to $0.3$ only 38 clusters shared top rank halo. These 38 clusters mapped to 19 halos.

Our approach is similar to that of \citet{Gerke:2005}, who introduced the concept of the largest joint member fraction to map clusters to halos.  
 However, that work uses a boolean measure of cluster membership.  The probabilistic approach of \RM makes the strength definition equivalent to the largest group fraction used in \citet{Gerke:2005}.  Note that \citet{redMaPPerII:2014} use a similar approach to match pairs of clusters derived from different search algorithms applied to the same SDSS data.

\section{Pairwise Velocity PDF: Halo Contributions to Spectroscopic Membership} \label{sec:Halos-spectroscopic}

The study of RMIV assessed the validity of \RM photometric membership probabilities by using spectroscopic redshifts.  That work models the line-of-sight velocity distribution of central--satellite pairs as a Gaussian distribution with zero mean and a dispersion that scales with cluster richness and, implicitly, with halo mass.  After removing projected pairs having larger than escape velocities, the PDF of the remaining normalized pairwise velocities is modeled as a Gaussian plus a uniform background.

We begin by demonstrating that the simulated galaxy sample displays similar characteristics to the observations.  Unlike the observations, our knowledge of the halo membership of each galaxy allows us to deconstruct the spectroscopic likelihood into distinct halo contributions.

\subsection{Constructing the velocity PDF of cluster central--satellite pairs} \label{sec:velocityPDF}

\begin{figure}
   \centering
   \includegraphics[width=0.45\textwidth]{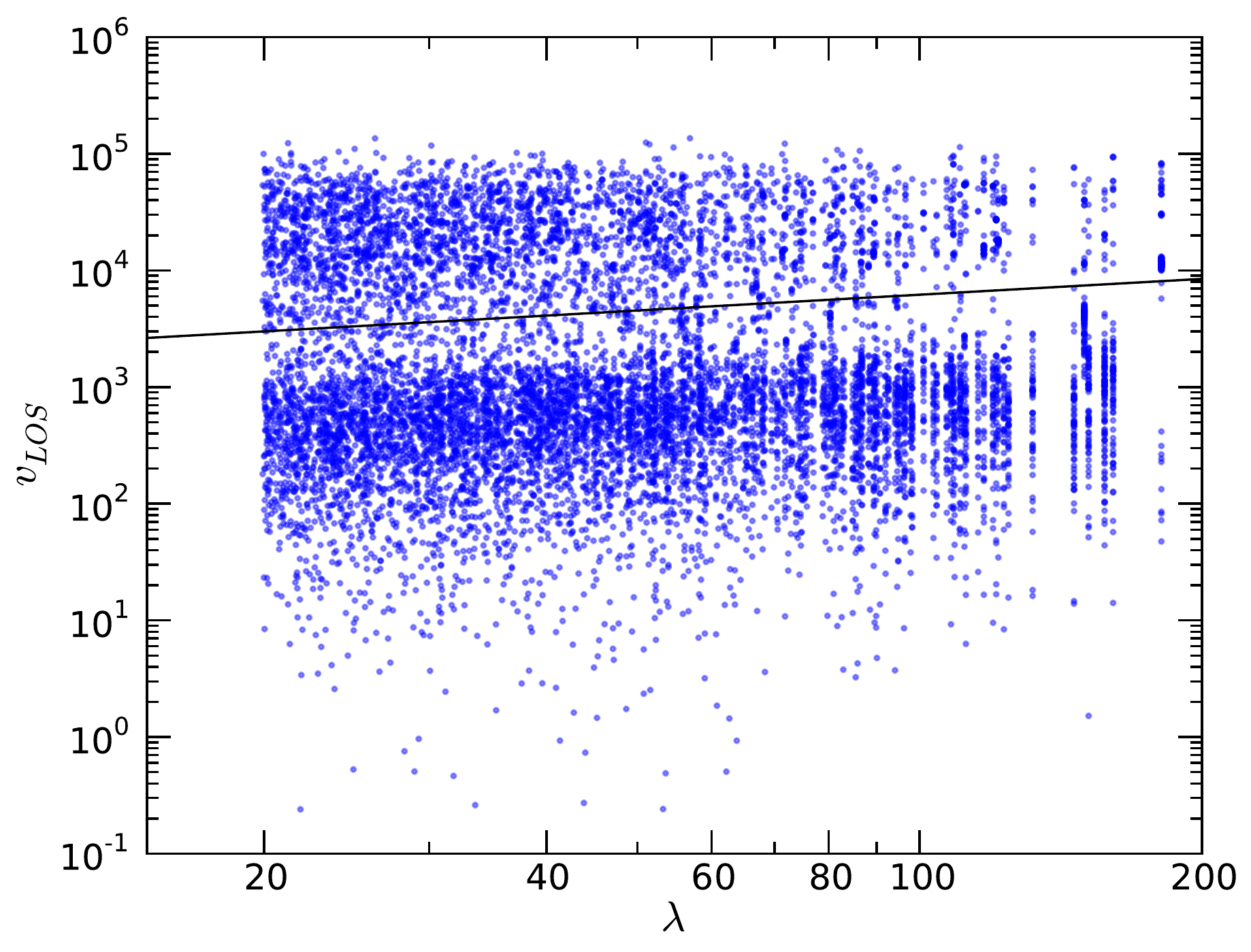}
 \caption{The line-of-sight magnitude of central--satellite pairwise velocities for all spectroscopic cluster members in the Aardvark simulation.  The line shows the cut applied applied to the SDSS sample by RMIV to separate cluster members (below) from projected contaminants (above).  We apply this cut to the Aardvark sample, eliminating $\sim 23 (25)\%$ of galaxy pairs from CEN (ALL) samples.  }
   \label{fig:velocityCut} 
\end{figure}

Using redshifts of cluster members in the spectroscopic samples described in section \ref{sec:specSamples}, we determine pairwise velocities of each cluster's satellite galaxies relative to its central galaxy 
\begin{equation}
   v = c \ \left(  \frac{z_{\rm gal} - z_{\rm cen}}{1+z_{\rm cen}} \right), 
   \label{eq:vpair} 
\end{equation}
where $c$ is the speed of light, and $z_{\rm gal}$ and $z_{\rm cen}$ are redshifts of satellite and central galaxies, respectively.  
The galaxy redshifts in the simulation are used with zero measurement error.  Recall that central galaxies of resolved halos are at rest with respect to their host halo.  

In Equation~(\ref{eq:vpair}), the central galaxy is defined by the \RM cluster-finding algorithm.  In the CEN sub-sample this is also the central galaxy of the matched halo. For clusters in the CEN sample with high strength, we expect the root mean square velocity to be an unbiased estimate of the dark matter velocity dispersion of the matched halo.  

Figure~\ref{fig:velocityCut} shows the distribution of pairwise velocity magnitudes against cluster richness for the ALL sample. The structure is very similar to that found by RMIV for the SDSS+GAMA spectroscopic data (see their Fig. 2), with a main component at low velocities, referred to as the signal, and a cloud of projected pairs lying at high velocities.  

We apply the RMIV velocity cut, shown by the line in Figure~\ref{fig:velocityCut}, to remove the projected contamination, eliminating $\sim 23 (25)\%$ of galaxy pairs in CEN (ALL) sample. 

As per RMIV, we model the velocity of a 
central-satellite pair as a random draw 
from a Gaussian distribution with a richness and redshift dependent velocity dispersion,  
$\sigma_v$, modeled via
\begin{equation} \label{eq:velocityDispersion}
   \sigma_{v} (\lambda,z_{cen}) = \sigma_{p} \left(\frac{1+z_{cen}}{1+z_p}\right)^{\beta}  \left(\frac{\lambda}{\lambda_p}\right)^{\alpha}
\end{equation}
where $\sigma_{p}$ is the characteristic dispersion at the pivot point, $\lambda_p=30$ and $z_p=0.2$ \footnote{In this work, the RMIV normalization is calculated using pivot richness, $\lambda_p=30$, and redshift, $z_p=0.2$, slightly different from the published RMIV pivot values.}, corresponding to the approximate median cluster richness and redshift of our sample, respectively.  

To incorporate non-physically associated pairs, a flat
velocity component is added to the distribution. The likelihood of the stacked velocity distribution is given by the following sum over pairwise velocities, $v_i$, 
\begin{equation} \label{eq:liklihoodModel}
  \mathcal{L} = \prod_{i=1}^{N_{\rm pair}} \left[ p \, G(v_{i},\sigma_{v}(\lambda,z)) + (1-p) \frac{1}{2 v_{max}} \right] ,
\end{equation}
where $G(v_{i},\sigma_{v}(\lambda,z))$ is a Gaussian of zero mean and width $\sigma_{v}(\lambda,z)$, and $p$, $\alpha$, $\beta$, and $\sigma_{p}$ are free parameters to be determined by maximizing the likelihood.  Each $v_i$ is the line-of-sight (LOS) satellite--central pair velocity, equation~(\ref{eq:vpair}), and the product is over all pairs in the sample.  

As we shall see in \S\ref{sec:Halos-contribution}, 
the fraction of pairs contained in the central Gaussian, given by the parameter $p$, is not the same as the fraction of cluster members contributed by the top-ranked halo.

\subsection{Velocity PDF analysis} \label{sec:CSresults}

\begin{figure*}
  \begin{subfigure}[b]{0.32\textwidth}
       \centering
	\includegraphics[width=\textwidth]{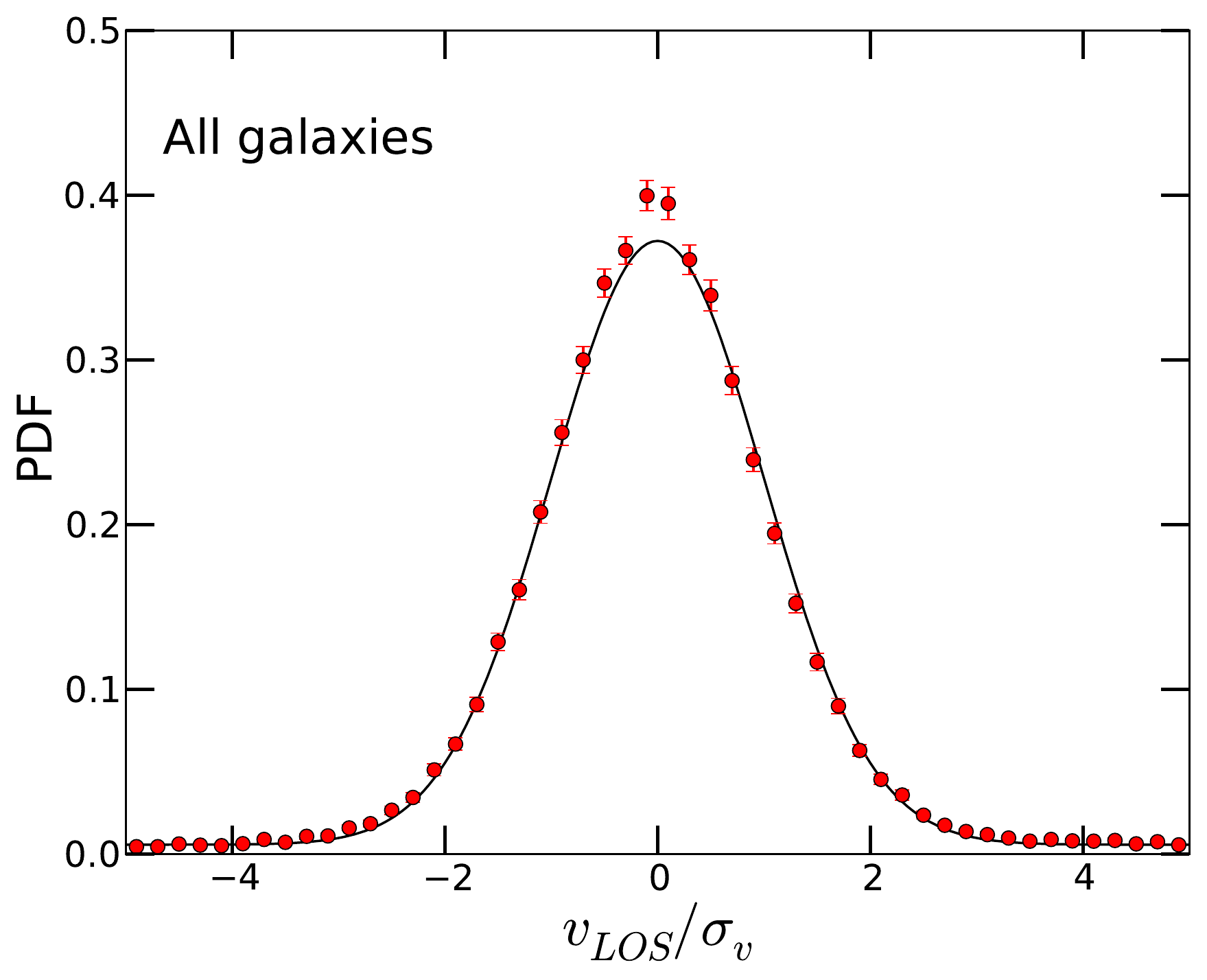}
   \end{subfigure}
   \begin{subfigure}[b]{0.32\textwidth}
       \centering
 	\includegraphics[width=\textwidth]{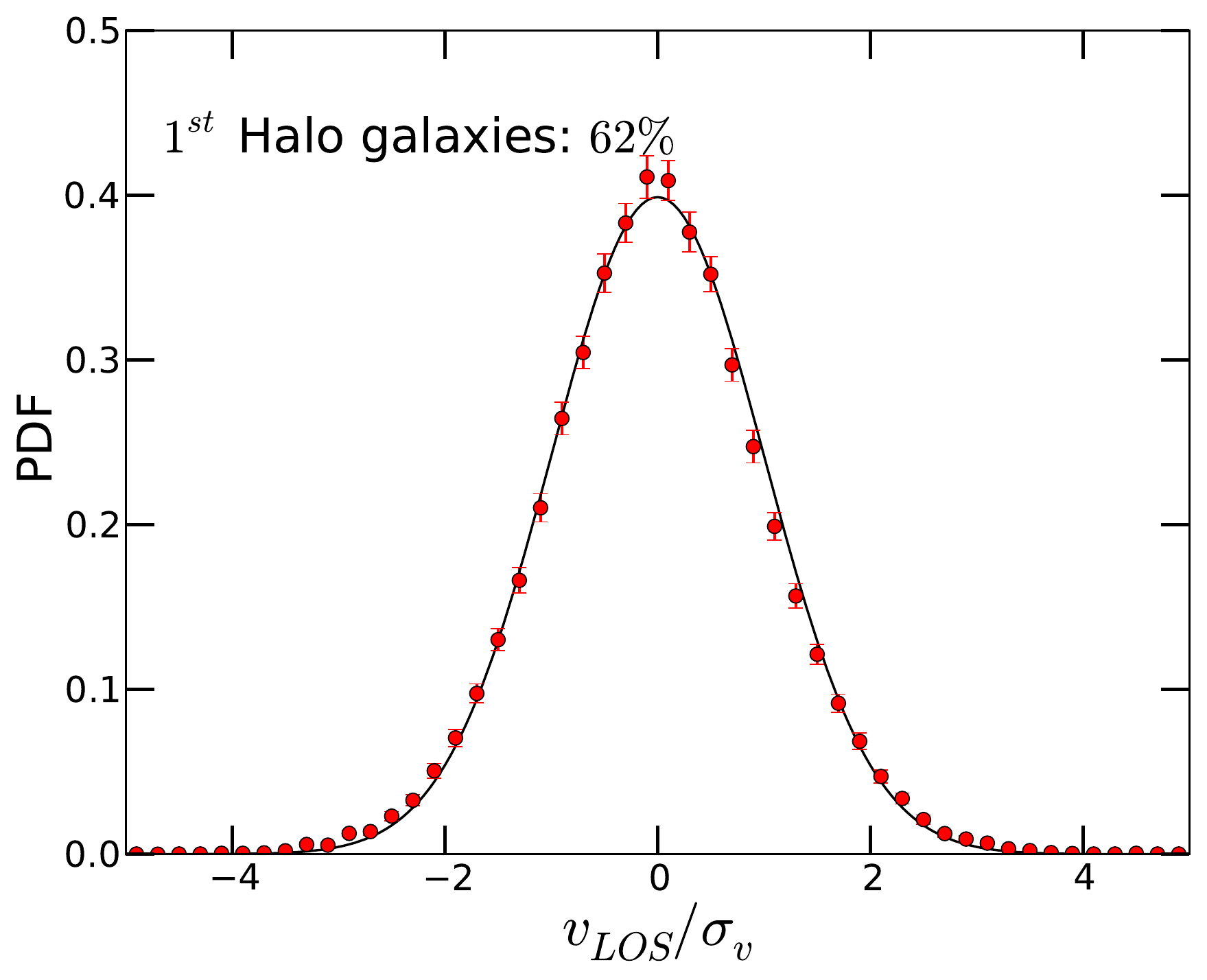}
   \end{subfigure}
   \begin{subfigure}[b]{0.32\textwidth}
       \centering
 	\includegraphics[width=\textwidth]{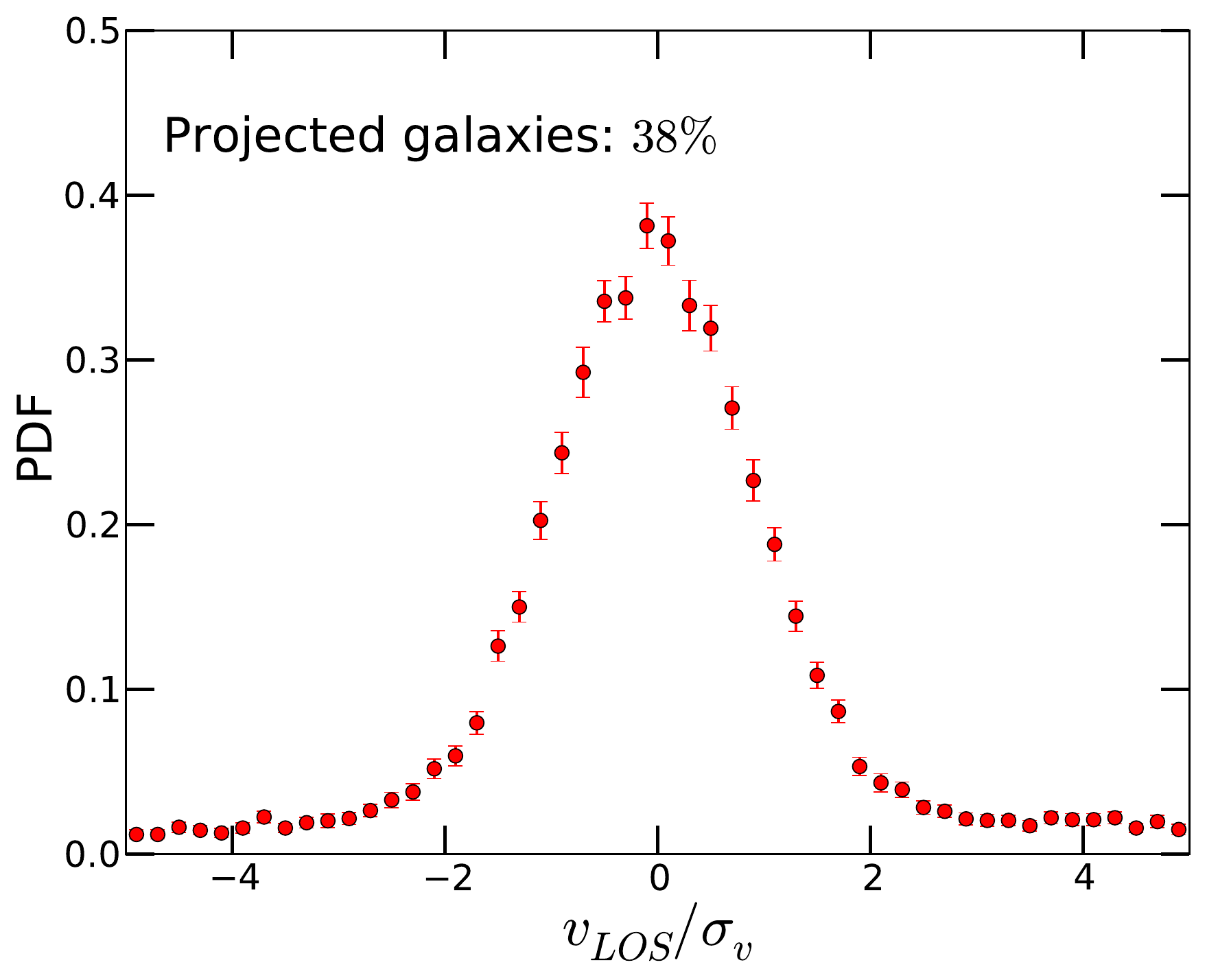}
   \end{subfigure}
   \caption{{\bf Left:} The PDF of LOS pairwise velocities, normalized according to equation \ref{eq:velocityDispersion}, for the correctly centered (CEN) sample of redMaPPer clusters in the Aardvark simulation.  The black line shows the best fit likelihood model, Equation~(\ref{eq:liklihoodModel}), with parameters given in Table~\ref{tab:specBestFit}.  {\bf Middle:}  Same as left but using only galaxy pairs in the matched (top-ranked) halo associated with each cluster.  The black line shows the likelihood model, equation \ref{eq:liklihoodModel}, but with $p=1$. Error bars are $2 \sigma$ based on bootstrap resampling. {\bf Right:}  Velocity PDF of galaxy pairs not belonging to the matched halo. }
   \label{fig:velocityDispersionClusters}
\end{figure*}

We maximize our likelihood to recover the scaling relation parameters between cluster richness and
velocity dispersion.  We assume flat priors on all parameters to calculate the posterior probability, and find the best-fit values given in Table~\ref{tab:specBestFit}. 

The left panel of Figure~\ref{fig:velocityDispersionClusters} shows the PDF of the pair velocities normalized by the expected velocity dispersion for the CEN cluster sample.
The structure of the full sample is similar.  We bootstrap the cluster sample to compute means and standard deviations of the PDF in 50 bins between $-5$ and $5$ in $v/\sigma_v$, shown as the points with error bars.  
The black line is a Gaussian of zero mean and unit variance plus the constant distribution, with amplitude given by the best fit model.

We find parameters that are similar to the RMIV fit to the SDSS redMaPPer sample. 
The CEN sample's Gaussian magnitude, $p=0.919 \pm 0.002$, and velocity--richness slope, $\alpha = 0.405 \pm 0.008$, are 
very similar to the SDSS values of $0.916 \pm 0.004$ and $0.44 \pm 0.02$, respectively.  The ALL sample has reduced magnitude, $p=0.885 \pm 0.002$ and a slightly shallower slope, $\alpha = 0.387 \pm 0.007$, differences that we discuss further in \S\ref{sec:miscentering} below.  

The velocity normalization, $\sigma_p$, is generally $\sim 10\%$ lower than the RMIV value.  As we discuss in section \ref{sec:massCalibration}, non-zero central galaxy velocities, satellite galaxy velocity bias, cosmology, and mis-centering frequency all play a role in setting the normalization. 

\begin{table*}
    \begin{center}
   \caption{Best fit parameters of the velocity dispersion model, equation~(\ref{eq:velocityDispersion}), using the likelihood, equation~(\ref{eq:liklihoodModel}) for the simulations (ALL, CEN, and Bolshoi), and the observational data of RMIV. Note that RMIV normalization is calculated at the pivot point, $\lambda_p=30$ and $z_p=0.2$, used in this paper. The Bolshoi simulation used only the $z=0$ simulation snapshot (Appendix~\ref{app:Bolshoi}) so cannot constrain $\beta$.  The quantity $\avg{f_{\rm h1}}$ is the mean fraction of spectroscopic cluster members contributed by the top-rank, matched halo.
} \label{tab:specBestFit}
   \begin{tabular}{ c c c c c c}
    \hline
      sample & $\sigma_p~[\kms]$ & p & $\alpha$ & $\beta$ & $\avg{f_{\rm h1}}$ \\ \hline
    ALL          & $585 \pm 2$ & $0.885 \pm 0.002$  & $0.387 \pm 0.007$ & $0.83 \pm 0.07$  &  0.58  \\ 
    CEN         & $547 \pm 2$ & $0.919 \pm 0.002$  & $0.405 \pm 0.008$ & $0.87 \pm 0.08$   &  0.62 \\ 
    Bolshoi  & $535 \pm 4$ & $0.884 \pm 0.003$ &  $0.295 \pm 0.010$  & -                           & 0.70 \\ 
    RMIV     & $598 \pm 6$ & $0.916 \pm 0.004$  & $0.435 \pm 0.020$  & $0.54 \pm 0.19$  & - \\ 
    \hline
    \end{tabular}
      \end{center}
\end{table*}

As an independent check that explores the sensitivity of these parameters to the galaxy assignment scheme, we repeat the analysis using measurements at known halo locations of the Bolshoi simulation catalogs of \citet{Hearin:2013}.  That work uses age distribution matching, a method for predicting how galaxies of magnitude $r$ and color $g-r$ occupy haloes, to populate halos with galaxies at redshift $z=0$.  
When using the galaxy catalog from the Bolshoi simulation, we rely on a $z=0$ snapshot rather than a properly constructed lightcone.  We note the \citet{Hearin:2013} catalog has only $g$ and $r$
data available, rather than the full 5-band photometry available in the SDSS and Aardvark.

Results of this exercise, details of which are given in Appendix~\ref{app:Bolshoi}, produce a velocity PDF of similar shape to the Aardvark CEN sub-sample. The best-fit parameters show a similar Gaussian magnitude, $p=0.89$, but a shallower slope, $\alpha = 0.30$, that reflects the steeper HOD slope in the Bolshoi galaxy catalog compared to the Aardvark galaxy catalog.

As found by RMIV, the best-fit model does not have an acceptable $\chi^2$, as reflected by the deviations seen in the left panel of Figure~\ref{fig:velocityDispersionClusters}.   
We show below that the deviations from the simple flat-plus-Gaussian model arise from galaxies lying along the line of sight in halos outside the matched halo.

\subsection{Halo-ranked contributions to the velocity PDF} \label{sec:Halos-contribution}
 
The cluster--halo membership network allows us to determine what fraction of pairs in the main Gaussian PDF component arise from the matched halo.  For the CEN sample, we find that, on average, $62 \%$ of galaxy pairs arise from within the matched halo.  For the full sample, the mean value decreases somewhat, to $58\%$.  For the Bolshoi catalog, in which all clusters are correctly centered by construction, the mean is somewhat larger, $70\%$. 
 
The middle panel of Figure~\ref{fig:velocityDispersionClusters} shows only the matched halo's contribution to the pairwise velocity PDF of the CEN sample.  As before, error bars are produced via bootstrap resampling of the cluster sample using $50$ bins between $-5$ and $5$ in $v/\sigma_v$. The black line shows a Gaussian with dispersion given by the best fit to the entire spectroscopic sample (left panel), listed in Table~\ref{tab:specBestFit}.  The principal 
difference with the left panel is that we force $p=1$, meaning no background component. While there exists moderate kurtosis in this distribution, the high velocity wings of the PDF are not well populated.  

The good match seen in the middle panel is important in that it indicates that the best-fit velocity derived from the  spectroscopic data set accurately recovers the velocity dispersion of the top-ranked halo.  This finding offers leverage for a mean dynamical mass estimate as a function of cluster richness that we explore in the next section.  
 
The right hand panel of Figure~\ref{fig:velocityDispersionClusters} shows the contribution from satellite galaxies outside of the matched halo.  Clearly, a constant background does not adequately capture this component, which is a sum over second and higher-ranked halos.  
For the CEN sample, an average of $38\%$ of spectroscopic pairs are not contributed by the top-ranked halo.  Of this total, an average of $10\%$ and $5\%$ come from the second- and third-ranked halo, respectively.  The remaining $23\%$ is contributed by fourth and higher ranked halos, with $12\%$ in unresolved halos below our mass resolution limit.

For the full sample (ALL), the overall non-matched halo fraction is slightly higher, $42\%$, with $12\%$ and $6\%$ arising from the second and third halo terms.   

Similar results have been found in prior simulation studies.  
Using a spectroscopic group finder based on a Voronoi-Delaunay tesselation, \citet{Gerke:2005} and \citet{Gerke:2012} find that $70\%$ of cluster galaxies truly belong to the matched host halo, on average. 
Though they use a completely different group finder algorithm, their conclusion regarding the level of interloper galaxies is consistent with the results of our spectroscopic analysis. 
In a different study, \citet{Mamon:2010}
finds the density of interloping dark matter particles in redshift space around massive halos takes the form of a constant component plus a quasi-Gaussian component, similar to the structure seen in the right panel of Figure~\ref{fig:velocityDispersionClusters}.


\section{Mass Estimation} 
\label{sec:massCalibration}

In this section we derive a scaling relation between total mass and optical richness by applying the virial velocity scaling of massive halos to the pairwise velocity dispersion model described above. We compare this stacked dynamical mass to that derived from membership matching to halos, and find excellent agreement with the log-mean matched mass at fixed richness.  

We begin by using the CEN sample to avoid uncertainties caused by mis-centering, then investigate mis-centering in \S\ref{sec:miscentering}.  

\subsection{Cluster mass from dark matter virial scaling} \label{sec:virialScaling}

The classical virial theorem balances the kinetic energy with (modulo surface terms) half the gravitational potential energy of a halo, thereby offering a scaling law between velocity dispersion and mass within an enclosed radius.  In a study of multiple, independent N-body and adiabatic hydrodynamic simulations, \citet[][hereafter, E08]{Evrard:2008} calibrated the dark matter virial relation. 

In that work, the one-dimensional velocity dispersion of a halo, $\sigma_h$, is defined in an orientation-averaged fashion using particles within $\rtwoh$.  The dispersion is measured with respect to the mean dark matter velocity within that radius. 

E08 showed that the halo velocity dispersion of the population follows a power-law form with approximately log-normal scatter, meaning the conditional probability, $P(\ln (\sigma_h) | M,z) = {\cal{N}} (\ln (\sigma_{\rm DM}(M,z)),0.046)$, where $\cal{N}$ denotes a normal distribution, $\sigma_{\rm DM}(M,z)$ is the log-mean velocity dispersion at fixed mass and redshift, and $0.046$ is the scatter in $\ln (\sigma_h)$ at fixed mass.  

\begin{figure*}
 \includegraphics[width=0.65\textwidth]{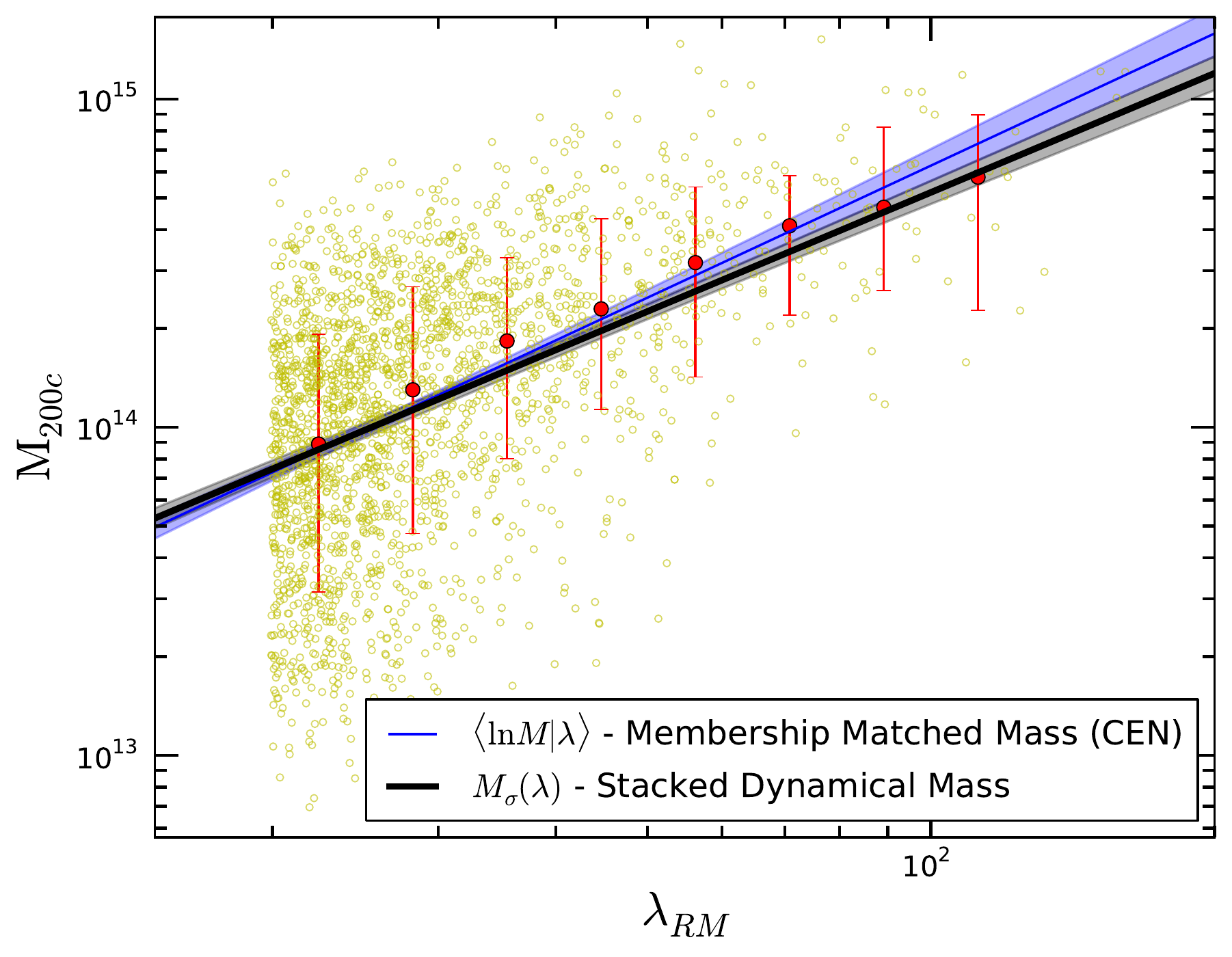}
   \caption{The mass--richness scaling relationship derived from application of the virial relation to stacked central satellite velocities, Equation~(\ref{eq:VTmass}), (solid black line) at redshift $0.2$ is compared to halo masses of correctly-centered redMaPPer clusters derived from galaxy membership matching in the redshift range $[0.1,0.3]$ (circles). The red dots with error bars show the median and $68\%$ inclusion region of matched halo mass in different richness bins. The blue line is the best fit to the membership-matched masses in this redshift range, with shaded region showing $95\%$ confidence uncertainties in this mean relation at redshift 0.2. }
   \label{fig:calMass} 
\end{figure*}

The log-mean velocity dispersion follows the scaling
\begin{equation}\label{eq:E08result}
\ln (\sigma_{\rm DM} (M_{200c},z) ) = \pi_\sigma + \alpha_\sigma  \ln (h(z)M_{200c} / 10^{15} ~\msol),  
\end{equation}
with amplitude $\pi_\sigma = \ln (1082.9 \pm 4.0)$ and slope $\alpha_\sigma = 0.3361\pm 0.0026$.  
Here, $h(z) = H(z)/100 \kms \mpc$ is the dimensionless Hubble parameter.  The ellipsoidal collapse model of \citet{Okoli:2015} offers a first-principles explanation of the form and parameter values of this calibration.  

At fixed mass, the distribution of velocity dispersion seen in the E08 simulation ensemble is very close to log-normal, with a modest tail to higher values driven by actively merging systems.  \citet{Saro:2013} show that the 1D LOS velocity dispersion has higher scatter compared to  angle-averaged velocity dispersion.  The normalization and slope of their scaling relation, found using sub-halos as galaxy tracers, are within $\lesssim 3\%$ of the E08 values.  

For a halo ensemble uniformly sampled in mass,  the inverse of the above scaling relation provides an unbiased estimate of the log-mean halo mass at fixed velocity dispersion, $P(\ln(M) | \ln(\sigma_h), z)$.  For samples drawn from the expected cosmic mass function, the log-mean mass will be biased low by approximately $5 \%$, as detailed in  \cite{Evrard:2014}. The magnitude of this correction is sub-dominant to systematic errors discussed below, so we choose to ignore it in this work. 

To estimate halo mass as a function of richness in the redMaPPer cluster population, we apply the inverse to the log-mean halo virial scaling relation found in E08, 
\begin{equation} \label{eq:VTmass}
\ln( h(z) M_\sigma(\lambda,z) /10^{15} \msol ) \  = \ 3 \ \ln \left( \frac{\sigma_v(\lambda, z)}{1083 \kms} \right) ,
\end{equation}
where $\sigma_v(\lambda, z)$ is the velocity dispersion scaling of central--satellite pairs analyzed in \S\ref{sec:Halos-spectroscopic} and the simple cubic power is consistent with the slope found in the E08 simulation ensemble.   

If intrinsic galaxy richness, $\lambda$, were a nearly perfect tracer of halo mass, and if cluster finders cleanly identified halo members, then the log-normal form of the PDF relating velocity to mass (or vice-versa) implies that the virial-scaled mass, $\ln(M_\sigma(\lambda,z))$, should accurately measure  the log-mean mass, $\langle \ln(M) | \lambda, z \rangle$, at fixed richness and redshift.  Introducing (log-normal) scatter in richness at fixed mass can produce shifts that depend on the local slope and curvature of the mass function as well as the covariance of $\lambda$ and $\sigma_h$ at fixed $M$ \citep{Evrard:2014}. We defer a more detailed examination of these issues to future work. 

Galaxy joint member matching provides an independent mass estimate for each cluster --- the matched halo mass --- that can used to assess the meaning of the stacked dynamical mass estimate, equation~(\ref{eq:VTmass}).  

Figure ~\ref{fig:calMass}, a key result of this work,  compares the mass scale inferred from the scaled velocity dispersion with membership matched masses for the CEN sample.
The thick black line shows the mass--richness scaling at redshift $0.2$ inferred from virial scaling, equation~\ref{eq:VTmass}, while the points show individual $M_{200c}$ values of matched halos for individual correctly-centered clusters of \RM richness, $\lambda$ within redshift range of $[0.1,0.3]$. The red dots with error bars show the median and $68\%$ inclusion region of matched halo mass in different richness bins.

The blue line and shaded blue region are the mean and $95\%$ uncertainty of a least-squares fit to the form, $\langle\ln M|\lambda, z\rangle = \pi_h + \alpha_h \log(\lambda/\lambda_p) + \beta_h \log((1+z)/(1+z_p))$.  We find parameters $\pi_h = \log(1.26 \pm 0.02 ~ [10^{14} ~ M_\odot]) $, $\alpha_h = 1.33 \pm 0.05 $ and $\beta_h = - 0.48 \pm 0.43 $.  The line is the $z=0.2$ relation while the shaded area shows combined uncertainties in the intercept and richness slope.   We find that the slope with redshift is consistent with zero with large uncertainties.

This virial scaling of stacked pairwise velocities is remarkably accurate in capturing the scaling with richness of the log-mean membership matched halo mass. Differences are less than $1 \%$ at the pivot point and within $\sim 5\%$ over a broad range in richness. Note that the E08 dark matter virial scaling is measured independently of the Aardvark simulation, so the level of agreement between the $\mtwoh$ and membership matched masses is a non-trivial result.

\begin{figure}
       \includegraphics[width=0.45\textwidth]{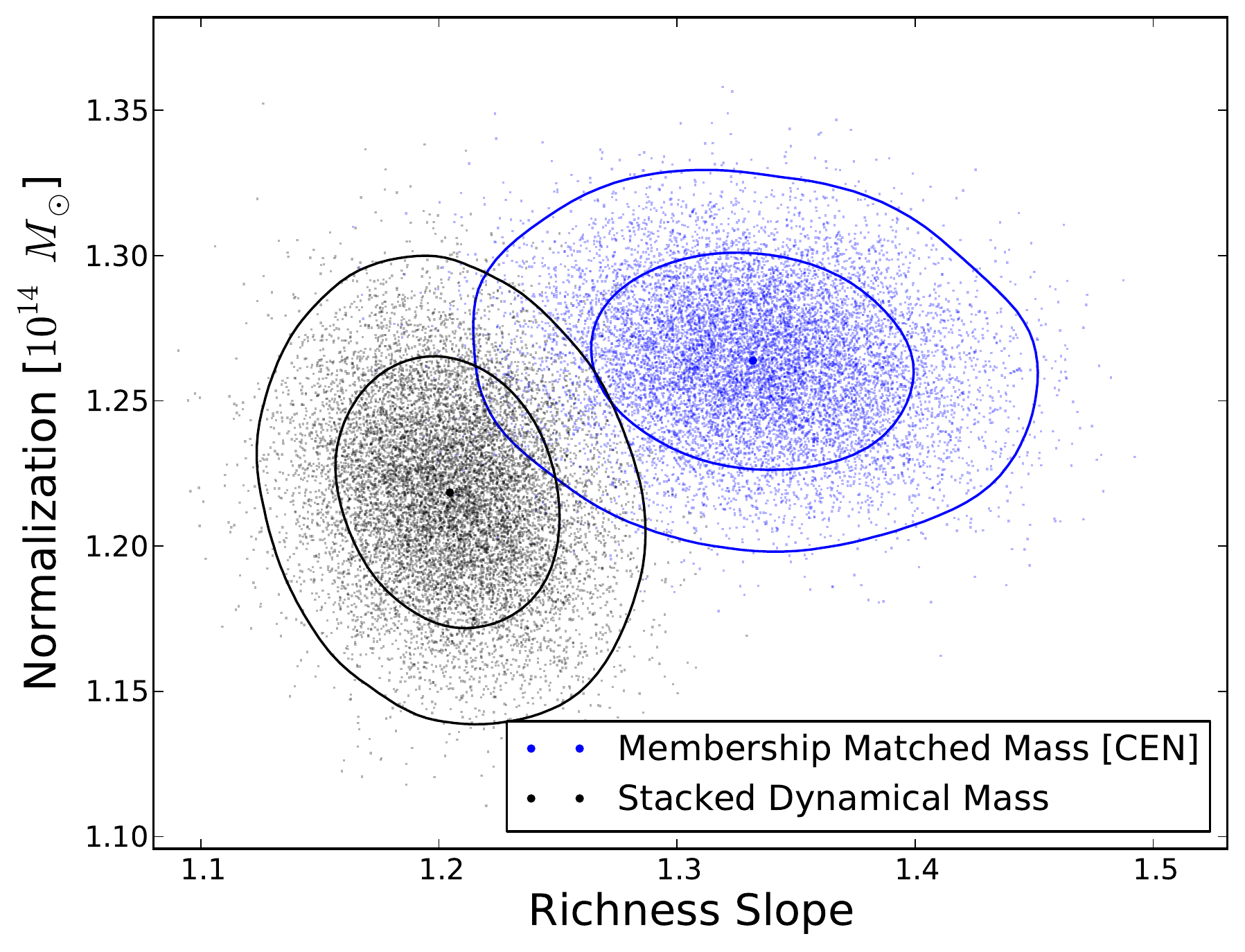}
   \caption{The normalization and slope of mass--richness scaling at redshift $0.2$ inferred from stacked dynamical masses (black contours) and membership matching in the redshift range $[0.1,0.3]$ (blue) for correctly-centered redMaPPer clusters. Contours show  $68\%$ and $95\%$  statistical uncertainties. }
   \label{fig:calMassContour} 
\end{figure}

Constraints on the slope and normalization of the mass--richness scalings for the CEN sample are compared in Figure~\ref{fig:calMassContour}.  
Black contours for the stacked dynamical mass include only statistical uncertainties in the constrained velocity parameters, not systematic errors discussed below.  
The blue contours are based on bootstrap resampling of membership matched halos within the redshift range $[0.18,0.22]$.  The normalizations at the $\lambda=30$ pivot are consistent, while the slopes are in mild tension at the level of 0.13 in their central value.  An ensemble of simulated samples would be useful to reduce the statistical error on the membership matched slope.

We turn next to discuss sources of systematic uncertainty before applying this method to derive a constraint the matched halo masses of RMIV clusters.

\subsection{Sources of Systematic Uncertainty}

The good agreement between stacked dynamical mass and membership-matched masses offers strong incentive to combine large photometric and spectroscopic galaxy samples to relate cluster richness to halo mass.  

Applying this method to survey data introduces several sources of systematic error that must be modeled.  The Aardvark synthetic sky realization is idealized in several respects; central galaxies are at rest with respect to their underlying halo and satellite galaxies trace the kinematics of the dark matter.  Also, the differences in stacked pairwise velocity model parameters for the CEN and ALL samples indicate that mis-centering plays an additional role.  In addition, variance in the velocity dispersion of clusters of fixed richness, reflective of the variance in matched halo mass, can introduce bias.  

The following sections address these issues in turn, finding that the first two are more important than the third. How satellite galaxies trace dark matter kinematics is the key source of systematic error.

\subsubsection{Central galaxy velocities and satellite galaxy velocity bias} \label{sec:velocityBias}

The degree to which galaxy velocities trace the kinematics of dark matter particles in halos is a central issue for virial mass calibration. 
By construction, the central galaxy is at rest with respect to its host halo in our simulations. In reality, central galaxies are measured to have a non-zero velocity dispersion with respect to their host clusters.  

In cases of actively merging systems the rest frame of a cluster is often difficult to define.  In the post-merger phase, the central galaxy will settle to the center of cluster due to dynamical friction on a timescale on the order of $1 \gyr$ \citep{White:1976, Bird:1994}, during which time the central galaxy will have a net velocity with respect to the full halo. 
Based on a sample of nearly 500 Abell clusters with 10 or more redshifts, \citet{Cozoil:2009} find that brightest cluster galaxies have velocities with root mean square magnitude $\sim 0.3 \sigma_{\rm cl}$, with $\sigma_{\rm cl}$ the line-of-sight velocity dispersion of the host cluster.  A similar ratio of $0.25$ is found by \citet{Lauer:2014} using 178 clusters with 50 or more member spectra.  \citet{Martel:2014} find a similar thermal motion for central galaxies in a sample of 18 massive halos extracted from a large cosmological, hydrodynamic simulation.  

Redshift-space distortion studies also support non-zero values for central galaxy velocities \citep{skibba:2011,Guo:2015,GuoII:2015}. If the central galaxy population has velocity dispersion scaling as some fraction, $\alpha_{\rm c}$, of the host halo dispersion, 
$\sigma_{\rm cen} = \alpha_{\rm c} \sigma_{\rm halo}$, then the central--satellite pairwise velocity normalization, $\sigma_p$, will be enhanced by a factor $(1+\alpha_{\rm c}^2)^{1/2} \simeq 1+ \alpha_{\rm c}^2/2$, the latter if $\alpha_{\rm c}$ is small compared to unity. Mass estimates derived from virial scaling will be boosted by a factor $(1+\alpha_{\rm c}^2)^{3/2} \simeq 1+ 3\alpha_{\rm c}^2/2$ relative to the case of cold centrals ($\alpha_c=0$).  These factors assume that the satellite galaxy velocities are unbiased with respect to the dark matter. 

The velocity dispersion of satellite galaxies relative to the halo rest frame may also biased  \citep{Carlberg:1994}, so that  
$\sigma_{\rm sat} \ = \ \alpha_{\rm s}  \ \sigma_{\rm DM}$,
where $\alpha_{\rm s}$ is the satellite galaxy velocity bias. 
The simulation study of \citet{Wu:2013} that combines N-body and hydrodynamic models indicates that $\alpha_{\rm s}$ lies near unity, with brighter galaxies tending to have values less than one and fainter galaxies slightly above unity, asymptotically reaching a value of $1.05$.  This pattern is not seen in the redshift-space distortion work of \citet{Guo:2015}, discussed below.

Let $\sigma_{p,0}$ be the normalization of the central--satellite pairwise velocity dispersion determined through the simulation analysis presented in \S\ref{sec:CSresults}.  Recall that the simulations are constructed to have $\alpha_{\rm c} = 0$ and $\alpha_{\rm s} = 1$.  Introducing uncorrelated central and satellite galaxy velocity biases modifies the pairwise velocity PDF normalization to 
\begin{equation} \label{eq:normVbias}
   \sigma_p \ = \ (\alpha_{\rm s}^2 + \alpha_{\rm c}^2)^{1/2} \sigma_{p,0} .
\end{equation}

If these effects alone are responsible for the normalization difference between the SDSS and Aardvark CEN samples (see Table~\ref{tab:specBestFit}), then we would require $(\alpha_{\rm s}^2 + \alpha_{\rm c}^2)^{1/2} = 1.13$.

\subsubsection{Cluster mis-centering} \label{sec:miscentering}

While the analysis of \S\ref{sec:CSresults} focused on the well-centered  subsample of clusters, the pairwise velocity PDF of the full sample has a similar form.  However, the fit parameters in Table~\ref{tab:specBestFit} indicate that the normalization of the full sample is enhanced, 585 (ALL) versus $547 \kms$ (CEN), and the slope $\alpha$ is slightly decreased.  
Because of the simulation limitations discussed in \S\ref{sec:simulation}, the mis-centered fraction of simulated \RM clusters in the ALL sample is larger than that of the SDSS sample.  Comparing to X-ray observations of a joint sample of more than 100 clusters, 
\citet{redMaPPerII:2014} find that $86 \pm 4 \%$ of high mass clusters are correctly centered on the X-ray counterpart.  This statistic is weighted 
toward higher richness values, $\lambda \sim 100$, but preliminary results of ongoing \RM sample analysis indicate that the full sample of $\lambda > 20$ \RM clusters has a similar fraction of well-centered clusters.

We exploit the differences in the CEN and ALL samples to estimate, using a weighted sampling approach, how velocity PDF parameters shift as the fraction of mis-centered clusters is varied.  

The ALL cluster sample contains both mis-centered and correctly centered clusters. Let $f_{\rm cen}$ be the fraction of ALL galaxy pairs lying in the latter (CEN) sample. Our approach is to simply create simulated central-satellite pairs drawn in proportion from the CEN and (ALL-CEN) cluster samples in order to achieve a desired $f_{\rm cen}$ value.  

Specifically, for a given $f_{\rm cen}$ value, we randomly draw without replacement a total of 10,000 galaxy pairs from these two cluster sub-populations in a way that satisfies the $f_{\rm cen}$ fraction.  We run the MCMC chains for these samples to find the best fit velocity PDF parameters for a total of 2000 realizations uniformly spanning $0.5 \le  f_{\rm cen} \le 1$.  

\begin{figure}
   \centering
    \includegraphics[width=0.45\textwidth]{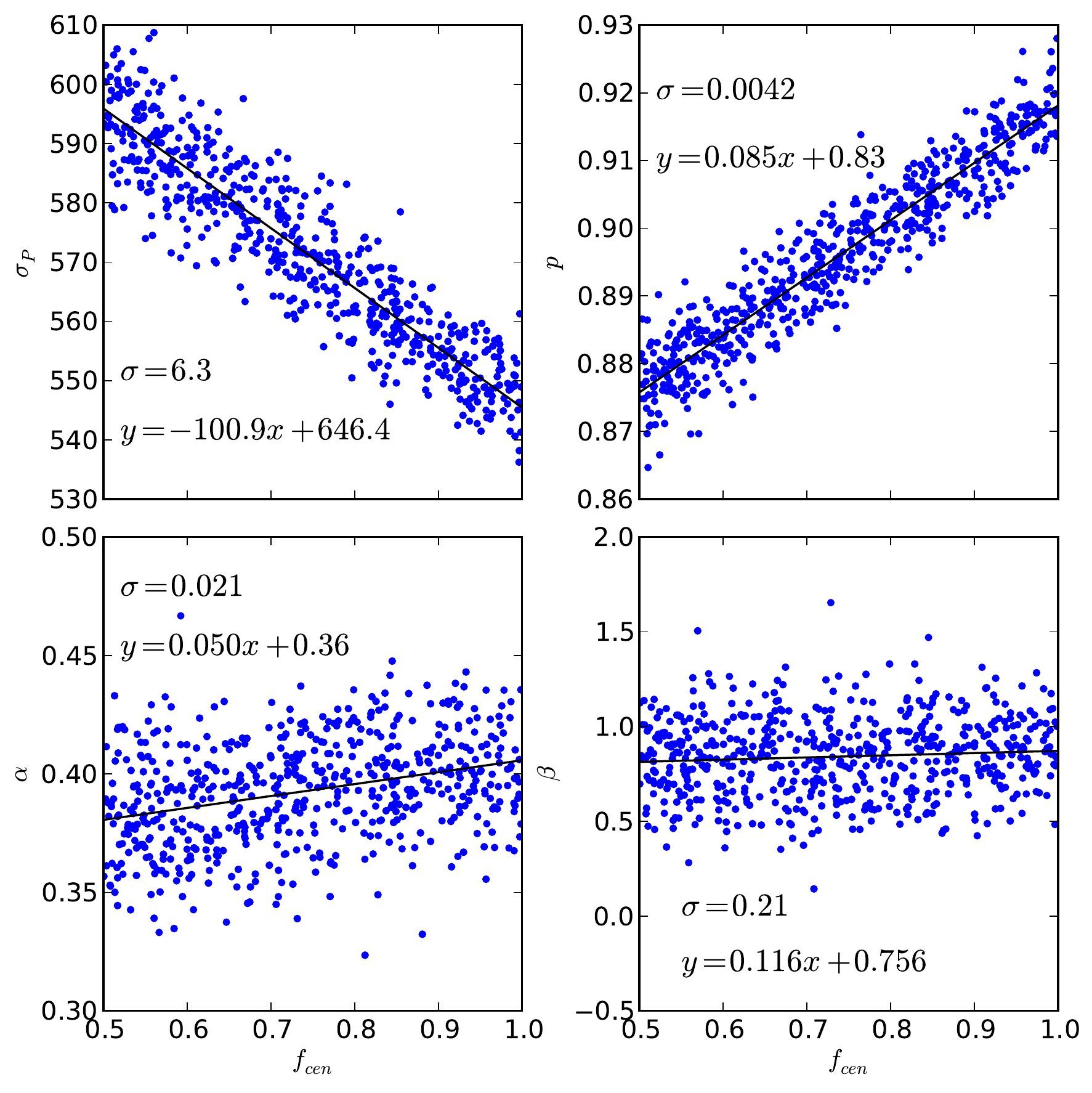}
 \caption{Sensitivity of the pairwise velocity PDF model to the fraction of correctly centered clusters, $f_{\rm cen}$.  Each point is derived from 10,000 galaxy pairs drawn randomly from the Aardvark CEN and (ALL-CEN) catalogs weighted to achieve the desired $f_{\rm cen}$.  The black lines show the best linear fit for each parameter, with the fit and standard deviation, $\sigma$, quoted in each panel. } 
   \label{fig:centeringIssue} 
\end{figure}

Figure~\ref{fig:centeringIssue} shows how the velocity PDF parameters change with correctly centered fraction, $f_{\rm cen}$.  
The black lines are the best linear fits as a function of $f_{\rm cen}$, with fit parameters and their root mean square deviations, $\sigma$, listed in the legend of each panel.  
 
As the fraction of mis-centered clusters increases (lower $f_{\rm cen}$ values), the velocity dispersion normalization, $\sigma_p$, increases while the slope, $\alpha$, and Gaussian amplitude, $p$, both decrease.  As expected, the limit of $f_{cen} = 1$ recovers parameters of the CEN catalog (see Table~\ref{tab:specBestFit}).

We use this behavior to correct for the effect of mis-centering on the RMIV pairwise velocity normalization.  
Assuming the fraction of correctly centered SDSS \RM clusters with $\lambda > 20$ to be $f_{\rm cen} = 0.85 \pm 0.05$\footnote{The value of $0.85 \pm 0.05$ is slightly more conservative than that published for higher richness clusters in \citet{Rozo:2014}.} leads to a $\sim 3\%$ normalization correction for correctly centered systems, 
\begin{equation}\label{eq:sigmaRMIVCC}
\sigma_{p,{\rm RMIV,CEN}} = 582 \pm 8 \ \kms. 
\end{equation}
We use this value to evaluate the mass scale of SDSS \RM clusters in \S\ref{sec:RMIVmass} below.  The mis-centering correction to the slope, $\alpha$, is smaller than $0.01$ and is not applied below.

\subsubsection{Velocity dispersion variance at fixed richness} \label{sec:scatterEffect}

The satellite--central velocity likelihood model employs a single Gaussian of width $\sigma_p(\lambda,z)$ at fixed richness, $\lambda$, but there is non-zero variance in velocity dispersion values of a fixed-$\lambda$ population that reflects the variance in matched halo mass.  Scatter in halo mass at fixed lambda is already incorporated into the simulations; the scatter in matched halo masses shown in Fig.~\ref{fig:calMass} is $0.85$ in $\ln M$.   We perform here an explicit test, independent of the simulated samples, to confirm that this scatter does not strongly affect the recovered velocity PDF parameters.  

We create ensembles of 10,000 galaxy pairs drawn from Gaussian distributions with dispersion values log-normally distributed about a scaling mean relation, $\sigma_p(\lambda,z)$, equation~\ref{eq:velocityDispersion} with variance $\sigma_{\ln \sigma}^2$. Sampling in $\lambda$ and redshift uniformly covers the observed ranges of $[20,200]$ and $[0.1,0.3]$, respectively.  We then perform the stacked velocity PDF analysis on each simulated pair ensemble.  

We find that the model parameters remain unbiased until $\sigma_{\ln \sigma} > 0.2$, after which the tails of the velocity distribution begin to affect the normalization $p$ at the one percent or greater level. The recovered values of $\sigma_p$ and $\alpha$, the key parameters involved in mass estimation, are unaffected up to values of $\sigma_{\ln \sigma} = 0.5$, or $1.5$ scatter in $\ln M$. This degree of mass scatter is larger than either the simulated or observed \citep{redMaPPerII:2014} values.  
Variance in host halo velocity dispersion at fixed richness is therefore a negligible source of systematic error in the velocity PDF modeling and resultant mass estimates.


\section{Stacked Dynamical Mass Scaling of SDSS redMaPPer Clusters }\label{sec:RMIVmass}

The above analysis indicates that the mass determined through virial scaling of the pairwise velocity PDF normalization offers an unbiased estimate of the log-mean mass of halos matched via joint gaalxy membership.  

We now turn to estimate the characteristic $M_{200c}$ mass scale of correctly centered \RM clusters as a function of richness $\lambda$ at the pivot redshift $z_p=0.2$. Recall from \S\ref{sec:velocityBias} and \ref{sec:miscentering} that the pairwise velocity normalization depends on the mis-centering frequency and the velocity bias of central and satellite galaxies.  We need to estimate the magnitudes of these effects, and their uncertainties, into our mass estimate. 

The normalization correction for mis-centering, assuming $f_{\rm cen} = 0.85 \pm 0.05$ for the SDSS \RM sample, is already incorporated into the correctly-centered estimate given in equation~(\ref{eq:sigmaRMIVCC}).   

To estimate the velocity dispersion of the underlying dark matter from the pairwise satellite--central galaxy measurements, we need to divide the latter by the quadrature sum of the respective velocity bias factors, 
\begin{equation}\label{eq:sigmapDM}
\sigma_{p,{\rm RMIV,DM}} = \frac{ \sigma_{p,{\rm RMIV,CEN}} }{(\alpha_{\rm s}^2 + \alpha_{\rm c}^2)^{1/2} }  .  
\end{equation}

The velocity bias of galaxies has been recently investigated by \citet{GuoII:2015,Guo:2015} using SDSS galaxy clustering measured both in projected separation and in redshift space. 
We employ the \citet{Guo:2015} estimates for the velocity bias factors of bright ($M_r \sim -21.5$, as appropriate for the bulk of the spectroscopic galaxies in this study) galaxies (see their Fig.~8) of $\alpha_{\rm c} = 0.30 \pm 0.05$ and $\alpha_{\rm s} = 1.05 \pm 0.08$.
Their central galaxy dispersion is in line with previous estimates based on explicit spectroscopy of cluster members \citep{Cozoil:2009,Lauer:2014} 
as well as with recent simulation expectations \citep{Martel:2014}.  There is more contention on the velocity bias of satellite galaxies --- 
values less than one have been measured in recent simulations for bright galaxies in massive halos \citep{Munari:2013, Old:2013, Wu:2013} -- but the we note that the $2\sigma$ range of $\alpha_s \in [0.89,1.21]$ admits values less than unity.  

These velocity bias estimates imply a correction factor,  $(\alpha_{\rm s}^2 + \alpha_{\rm c}^2)^{-1/2} = 0.92 \pm 0.07$, which leads to the dark matter velocity dispersion at the pivot richness and redshift of 
\begin{equation}\label{eq:pivotVdisp}
\sigma_{p,{\rm RMIV,DM}} = 535 \pm 41 \ \kms.  
\end{equation}
Note that the uncertainty in this velocity is dominated by systematic error in the velocity bias estimate.  

Finally, using this value in equation~(\ref{eq:VTmass}), we obtain an estimate of the log-mean mass of \RM clusters at the pivot richness and redshift of 
\begin{equation}\label{eq:RMIVmass}
     M_\sigma(\lambda_p=30,z_p=0.2) = (1.56 \pm 0.35) \times 10^{14}\msol,
\end{equation}
where to infer above mass scale we assume a $\Lambda$CDM cosmology with $\Omega_{m} = 0.3$, $\Omega_{\Lambda} = 0.7$, and $h(z=0)=0.7$.  

The scaling of the pairwise velocity normalization, $\sigma_p(\lambda,z)$, determines how the mean dynamical mass, $M_\sigma(\lambda,z)$, scales with richness and redshift. Because of the relatively weak constraint on the redshift scaling behavior of the SDSS cluster sample velocities, we defer analysis of redshift evolution to a later study and concentrate here on the behavior with richness at the pivot redshift of $0.2$.  The simulations indicate the the mean dynamical mass, $M_\sigma(\lambda,z)$, matches the log-mean membership matched mass at the pivot richness, but as shown in Figure~\ref{fig:calMassContour}, the best-fit slope of log-mean mass with richness differs by 0.10 from the slope of $M_\sigma(\lambda)$.  We therefore include this difference as a systematic error term when quoting the slope.

The result is an estimate for the log-mean membership matched mass of the SDSS \RM sample at redshift $0.2$ of 
\begin{equation}
  \avg{\ln(M_{200c} /10^{14}\msol)|\lambda,z_p=0.2} = \pi + \alpha_m \, \ln(\lambda / 30)
\end{equation} 
with normalization $\pi = 0.44 \pm 0.22$ and slope $\alpha_m = 1.31 \pm 0.06_{stat} \pm 0.13_{sys}$.

Of the 22\% error in the derived mass normalization, 21.5\% arises from systematic uncertainty in the velocity bias terms, particularly that of satellite galaxies.  Mis-centering contributes $2.6\%$, and statistical uncertainties from the stacked pairwise velocity and virial calibration parameters are $3.2\%$.  The error in $\ln(M)$ is essentially triple the uncertainty in $\ln(\alpha_s)$.  As a result, achieving ten percent error in mean mass would require knowing $\alpha_s$ to a fractional accuracy of $\sim 0.03$. 
It remains to be seen whether future spectroscopic campaigns, coupled with improved hydrodynamic simulations of galaxy formation in massive halos to pin down systematic errors, can achieve this level of precision.


\section{Conclusion} \label{sec:conclusion}

Using galaxy catalogs derived from large N-body simulations, we study the mapping 
of galaxy clusters identified in sky-photometry space to the underlying real-space population of halos through membership matching.  
We measure membership strength, defined as the fraction of a cluster's richness contributed by a given halo, and 
build bipartite graphs linking clusters to halos with strength-weighted edges.  The matched halo of a cluster maximizes this strength.  
  
We then study pairwise velocities, and derived masses, from stacked spectroscopic analysis of clusters patterned after the 
spectroscopic analysis of SDSS redMaPPer clusters developed by RMIV.  The structure in the simulated data is similar to that of the observations, with galaxy pairwise velocities having a main Gaussian provisionally identified as cluster members.  We employ a sub-sample of correctly centered clusters --- those for which the central cluster galaxy is also the central galaxy of the matched halo --- as well as studying the full simulated cluster sample.  

We then use our findings to estimate the log-mean, membership-matched mass of SDSS \RM clusters at $z=0.2$.  Our detailed results are as follows.

\begin{itemize}
\item The pairwise velocity PDF has a main component that is reasonably well modeled as a Gaussian, with richness and redshift dependent width, plus a constant.  Decomposing this main  component into halo contributions, we find that the top-ranked, matched halo contributes an average of $62 \%$ ($58\%$) of pairs in the correctly centered (full) cluster samples.  The second-ranked halo contributes $\sim 10\%$, the third $\sim 5 \%$, and the remainder contribute $\sim 20\%$, in the mean.  The projected component, consisting of all galaxy pairs not contributed by the top-ranked matched halo, has a pairwise velocity PDF described roughly by a Gaussian plus constant form.

\item Converting the velocity dispersion--richness relation to a mass--richness relation using the dark matter virial relation calibrated by independent simulations  \citep{Evrard:2008}, we find this stacked dynamical mass recovers, to within a few percent, the log-mean mass determined from membership matching between clusters and halos. 

\item We model effects of cluster mis-centering and galaxy velocity bias in order to correct the measured \RM cluster velocity dispersion to reflect that of correctly centered, dark matter halos.  Using central and satellite velocity bias parameters $\alpha_c=0.30 \pm 0.05$ and $\alpha_c=1.05 \pm 0.08$, respectively \citep{Guo:2015}, we infer a log-mean matched halo mass of $M_{200,p} = (1.56 \pm 0.35) \times 10^{14} \msol$ at the pivot richness, $\lambda_p = 30$, and redshift $z_p = 0.2$, and a slope with richness of $1.31 \pm 0.06_{stat} \pm 0.13_{sys} $ for SDSS \RM clusters.
\end{itemize}

Kinematic biases of central and, especially, satellite galaxies, are the dominant source of systematic error.  Further work is needed, both empirically and through hydrodynamic simulations, to better constrain the relationship between galaxy velocities and dark matter. One possible approach is to invert the analysis presented here; comparing the stacked dynamical masses with stacked weak lensing masses of the same sample with the aim of constraining velocity bias.

\section*{Acknowledgment}

We acknowledge support from the DOE Office of Science grant DE-SC0007859.  RHW received support from the National Science Foundation under NSF-AST-1211838.  We acknowledge our collaborators Matthew Becker, Michael Busha, Brandon Erickson, and Andrey Kravtsov for their contributions to creating the simulations used here. We used the affine-invariant MCMC sampler package, \texttt{emcee}\footnote{\url{https://github.com/dfm/emcee}} \citep{Foreman-Mackey:2013}, to perform MCMC analysis.

\bibliographystyle{mn2e}
\bibliography{mybib}


\appendix

\section{Spectroscopic membership in the Bolshoi simulation} \label{app:Bolshoi}

We provide here a test of the sensitivity of the pairwise velocity PDF to the galaxy assignment method based on the Bolshoi simulation catalog of \citet{Hearin:2013}.  

\citet{Hearin:2013} use age distribution matching, an empirical method for modeling how galaxies occupy haloes as a function of luminosity and colour. The method relates galaxy luminosity to sub-halo size, and assigns color to formation epoch at fixed size, with older systems being redder. Table~\ref{tab:HODtable} compares the HOD properties of the Aardvark and Bolshoi catalogs.  The slope and intrinsic scatter for a luminosity cut, $M_r -5\log h = -19$, are given for halo samples limited in mass above $M_{200c} = 2 \times 10^{14} \hinv \msol$ (Aardvark) and $M_{vir} = 10^{14} \hinv \msol$ (Bolshoi).  The Aardvark HOD is shallower and has larger variance than that of Bolshoi.    

\begin{table}
    \begin{center}
    \begin{tabular}{ c c c c}
    \hline
      Simulation & $z$ & Slope & Scatter \\ \hline
    Aardvark     & 0.15 & 0.75  & 0.33  \\  
    Aardvark     & 0.25 & 0.79  & 0.28  \\  
    Bolshoi       & 0  &  1.0   & 0.21  \\ 
    \hline
    \end{tabular}   
    \caption{Slope and log-normal scatter of intrinsic richness, $\lambda_{int}$, versus mass,  $M_{200c}$, for the Aardvark and Bolshoi \citep{Hearin:2013} simulations applying a luminosity threshold of $M_r -5\log h = -19$.  } \label{tab:HODtable}
    \end{center}
\end{table}

Despite these differences in intrinsic galaxy population, 
we find that Bolshoi and Aardvark simulation both produce similar pairwise velocity PDF structure.  For the Bolshoi analysis, we place the $z=0$ catalog at an effective redshift 0.1, then use cylinders of length $120 \hinv\mpc$ (the size of the Bolshoi simulation) centered on known halo locations to extract projected galaxies around the position of the central galaxy. The $120 \hinv\mpc$ comoving length is equivalent to $\sim 0.03$ redshift shells at a central redshift of 0.1.

We measure \RM richness and generate LOS velocity pairs based on \RM cluster membership.  There is no redshift evolution in color because the Bolshoi redshift is fixed.  

\begin{figure*}
  \begin{subfigure}[b]{0.45\textwidth}
       \centering
       \includegraphics[width=\textwidth]{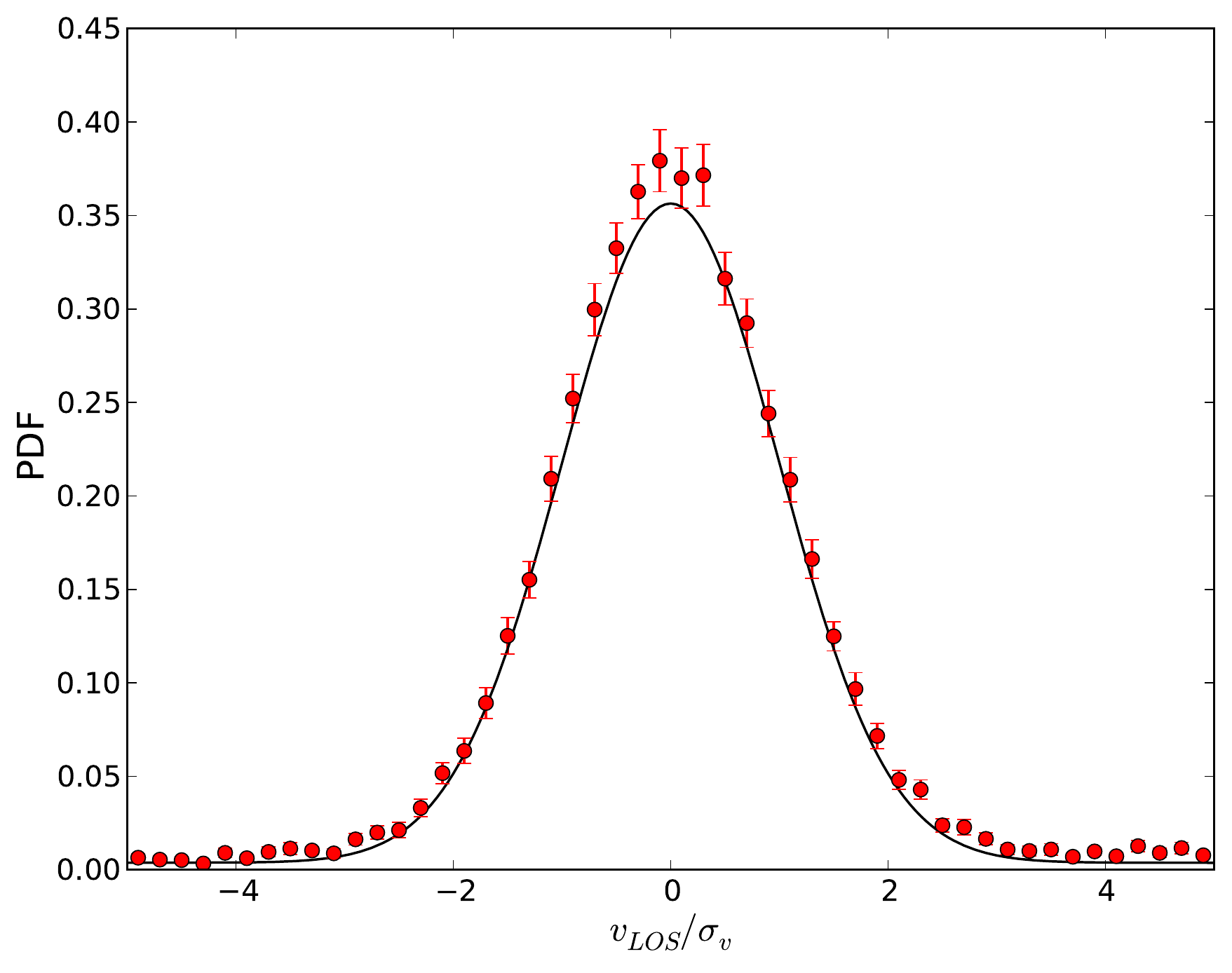}
   \end{subfigure}
   \begin{subfigure}[b]{0.45\textwidth}
       \centering
       \includegraphics[width=\textwidth]{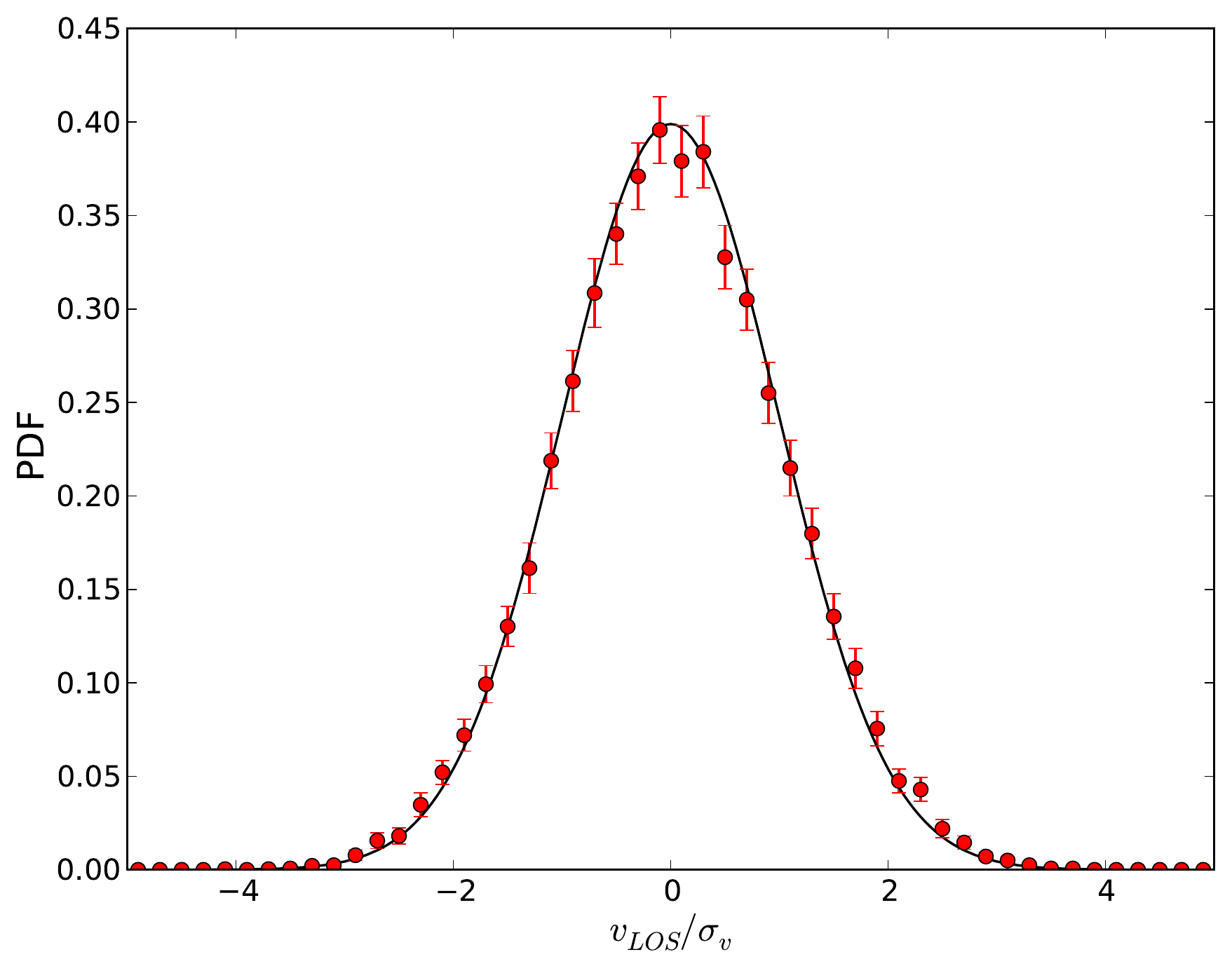}
   \end{subfigure}  \caption{The left panel shows the PDF of spectroscopic sample galaxy members' LOS velocity in the Bolshoi simulation by normalizing the velocity according to equation \ref{eq:velocityDispersion} and setting the redshift evolution to zero. The black line shows the best fit for our likelihood model equation \ref{eq:liklihoodModel}. The right panel shows similar exercise for the target halo galaxy members in the Bolshoi simulation. The black line shows the best fit for our likelihood model equation \ref{eq:liklihoodModel} assuming $p=1$. Error bars are two sigma using bootstrap method.}
   \label{fig:velocityDispersionClustersBolshoi}
\end{figure*}

Figure~\ref{fig:velocityDispersionClustersBolshoi} shows the normalized, pairwise velocity PDF's for all members (left panel) and using only member of the target halo (right panel).  These panels are the equivalent of the left and middle panels of Figure~\ref{fig:velocityDispersionClusters}.  
The results are generally consistent with the Aardvark simulation and the observational data.  Differences in the fit parameters listed in Table~\ref{tab:specBestFit} reflect differences in the intrinsic HODs of the two simulations.  

We find a Gaussian component amplitude of $p=0.884$, only slightly lower than the values of the Aardvark CEN sample and the observational data.  Using the membership definition of \citet{Hearin:2013} (used to calculate the intrinsic richness of the halo), we find that $70\%$ of spectroscopic member galaxies belong to the target halo.  The shallower slope $\alpha$ reflects the steeper intrinsic HOD slope of Bolshoi.  The larger strength, $S_{\rm max}$, of the matched halo membership contribution to the main Gaussian component may be due to the smaller HOD scatter as well as the more limited treatment of projected contamination in that simulation.

\label{lastpage}

\end{document}